\pdfoutput=1
\RequirePackage{ifpdf}
\ifpdf 
\documentclass[pdftex]{sigma}
\else
\documentclass{sigma}
\fi

\usepackage{amsfonts}
\usepackage{lipsum}

\usepackage{enumitem}

\usepackage{environ}

\newcommand\numbers{\textup{(\arabic*)}}
\newcommand\ZZ{\mathbb{Z}}
\newcommand\RR{\mathbb{R}}
\newcommand\CC{\mathbb{C}}
\newcommand\DD{\mathbb{D}}
\newcommand\cA{\mathcal{A}}
\newcommand\cO{\mathcal{O}}
\DeclareMathOperator\id{id}
\DeclareMathOperator\gr{gr}
\DeclareMathOperator\SL{SL}
\DeclareMathOperator\Res{Res}
\newcommand\bP{\textbf{P}}
\newcommand\rn{\frac{2\pi {\rm i}}{n}}
\newcommand\trn{\tfrac{2\pi {\rm i}}{n}}
\renewcommand\d{\mathrm{d}}
\let\Re\relax
\DeclareMathOperator\Re{Re}
\let\Im\relax
\DeclareMathOperator\Im{Im}
\newcommand\acts{\cdot}
\newcommand\eps{\varepsilon}

\newcommand\R{\bar{R}}
\renewcommand\S{\bar{S}}

\newcommand\half{\ensuremath{\frac{1}{2}}}
\newcommand\thalf{\ensuremath{\tfrac12}}

\newcommand{\mc}{\mathcal}
\newcommand{\ZN}{\mathbb{Z}}
\newcommand{\ovl}{\overline}
\newcommand{\Gr}{\operatorname{Gr}}
\newcommand{\PGL}{\operatorname{PGL}}

\numberwithin{equation}{section}

\newtheorem{Theorem}{Theorem}[section]
\newtheorem*{Theorem*}{Theorem}
\newtheorem{Corollary}[Theorem]{Corollary}
\newtheorem{Lemma}[Theorem]{Lemma}
\newtheorem{Proposition}[Theorem]{Proposition}

\theoremstyle{definition}
\newtheorem{Definition}[Theorem]{Definition}
\newtheorem{Remark}[Theorem]{Remark}

\begin{document}

\allowdisplaybreaks

\newcommand{\arXivNumber}{2507.19447}

\renewcommand{\PaperNumber}{033}

\FirstPageHeading

\ShortArticleName{Positive Traces on Certain ${\rm SL}(2)$ Coulomb Branches}

\ArticleName{Positive Traces on Certain $\boldsymbol{{\rm SL}(2)}$ Coulomb Branches}

\Author{Daniil KLYUEV~$^{\rm a}$ and Joseph VULAKH~$^{\rm b}$}

\AuthorNameForHeading{D.~Klyuev and J.~Vulakh}

\Address{$^{\rm a)}$~Department of Mathematics, Northwestern University, USA}
\EmailD{\mail{klyuev@northwestern.edu}}

\Address{$^{\rm b)}$~Department of Mathematics, Massachusetts Institute of Technology, USA}
\EmailD{\mail{jvulakh@mit.edu}}

\ArticleDates{Received July 28, 2025, in final form March 15, 2026; Published online April 10, 2026}

\Abstract{For a noncommutative algebra $\mathcal{A}$ and an antilinear automorphism $\rho$ of $\mathcal{A}$, there is a notion of a positive trace. When we have a three-dimensional $\mathcal{N}=4$ gauge theory or four-dimensional $\mathcal{N}=2$ gauge theory compactified on a circle, classification of positive traces on its Coulomb branch $\mathcal{A}$ can give a better understanding of this theory. We classify positive traces on $\mathcal{A}$ in two cases. The first case is when $\mathcal{A}$ is a quantization of a Kleinian singularity of type $D$, with certain restriction on the quantization parameter. The second case is when \smash{$\mathcal{A}=K^{{\rm SL}(2,\mathbb{C}[[t]])\rtimes \mathbb{C}_q^{\times}}({\rm Gr}_{{\rm PGL}_2})$} is an algebra containing $K$-theoretic Coulomb branches of pure ${\rm SL}(2)$ and ${\rm PGL}(2)$ gauge theories.}


\Classification{53D50; 81R10; 14D21}

\section{Introduction}
Let $\mc{A}$ be an algebra with an automorphism $g$. All algebras will be associative, unital and over~$\CC$. A {\it twisted trace} is a map $T\colon\mc{A}\to\CC$ such that $T(ab)=T(bg(a))$ for all $a,b\in\mc{A}$.

Let $\rho$ be an antilinear automorphism of $\mc{A}$ and $g=\rho^2$. A $g$-twisted trace $T$ is called {\it positive} if~${T(a\rho(a))>0}$ for all nonzero $a\in\mc{A}$. In this case, $(a,b)=T(a\rho(b))$ is a positive definite Hermitian form on $\mc{A}$ satisfying $(a\rho(b),c)=(a,bc)$.

In the case when $\mc{A}$ is filtered, Etingof and Stryker~\cite{etingof2020short} proved that twisted traces satisfying a certain nondegeneracy condition correspond to short nondegenerate star-products on $\gr\mc{A}$ introduced by Beem, Peelaers and Rastelli~\cite{BPR}. Positive traces correspond to unitary short star-products. In the case when $\mc{A}$ comes from a superconformal field theory, there is a certain unitary short star-product on $\gr\mc{A}$ that reflects the properties of the corresponding field theory. In some cases, a positive trace is unique up to scaling, which allows us to compute the corresponding unitary short star-product.

Another physical motivation comes from the work of Gaiotto~\cite{GaiottoSphere} and Gaiotto and Tesch\-ner~\cite{GT}. When $\mc{A}$ is a quantized Higgs/Coulomb branch of a superconformal field theory, there is a certain physically motivated choice of a conjugation $\rho_{\text{sph}}$ and a trace $T_{\text{sph}}$. We call $T_{\text{sph}}$ {\it sphere trace.} Physicists expect $T_{\text{sph}}$ to be positive and it would be useful to have a mathematical proof of that.
The sphere trace was introduced by Gaiotto and Okazaki~\cite{GO}, where they proposed a~formula for the sphere correlation functions of Coulomb/Higgs operators in terms of the sphere trace.

The algebra $\mc{A}$ does not have to be filtered, and Gaiotto and Teschner consider $K$-theoretic Coulomb branches. An advantage of $K$-theoretic Coulomb branches is that $\rho_{\text{sph}}$-positive trace should always be unique up to scaling. A mathematical proof of that is much more complicated than just proving positivity of $T_{\text{sph}}$ and the uniqueness statement is very useful for the following reason.

Suppose that $\mc{A}$ indeed has a unique $\rho_{\text{shp}}$-positive trace $T_{\text{sph}}$. In this case, we canonically produce a Hilbert space $\mc{H}$ with an action of $\mc{A}\times\mc{A}^{{\rm op}}$ as follows. The space $\mc{H}$ be a completion of $\mc{A}$ with respect to $(a,b)=T_{\text{sph}}(a\rho_{\text{sph}}(b))$ and $\mc{A}$ acts on $\mc{H}$ by unbounded operators of left and right multiplication $L_a$, $R_a$. The pair $(\mc{A},\mc{H})$ is useful for understanding the superconformal field theory that $\mc{A}$ came from. See \cite[Section~1.3]{GT} for the discussion of class $\mc{S}$ theories and the uniqueness of the trace. The same motivation applies to $K$-theoretic Coulomb branches of not necessarily conformal supersymmetric gauge theories; see \cite[Section~2.2]{GT} for the discussion of how do non-conformal theories fit into their picture.

In this article, we classify positive traces in the following two cases, described in more details below. The first is when $\mc{A}$ is a deformation of a Kleinian singularity of type $D$ satisfying a~certain condition; we study this case in Sections~\ref{sec:prelim}--\ref{sec:pos}. The second is when $\mc{A}$ is an algebra containing quantized $K$-theoretic Coulomb branches of pure $\SL(2)$ and $\PGL(2)$ gauge theories; we study this case in Sections~\ref{SecAlgebra}--\ref{SecPositiveTraces}.
In particular, our results imply that the sphere trace $T_{\text{sph}}$ is positive when $\mc{A}$ is the quantized $K$-theoretic Coulomb branch of pure $\SL(2)$ or $\PGL(2)$ gauge theory.

The first case we study deals with Kleinian singularities.
A Kleinian singularity is a quotient~$\CC^2 / G$,
where $G$ is a finite subgroup
of the special linear group $\SL_2(\CC)$.
The Kleinian singularity $\CC^2 / G$
has as its coordinate ring
the commutative graded algebra $A = \CC[u, v]^G$
of $G$-invariant polynomials in two variables.

Finite subgroups of $\SL(2,\CC)$ are classified up to conjugation by simply-laced Dynkin diagrams. The $A_{n-1}$ diagram corresponds to the cyclic subgroup generated by
\[
 \begin{pmatrix}
 {\rm e}^{\frac{2\pi {\rm i}}{n}} & 0\\
 0 & {\rm e}^{-\frac{2\pi {\rm i}}{n}}
 \end{pmatrix}.
\]
The $D_{n-1}$ diagram corresponds to the group generated by
\[
 \begin{pmatrix}
 {\rm e}^{\frac{\pi {\rm i}}{n}} & 0\\
 0 & {\rm e}^{-\frac{\pi {\rm i}}{n}}
 \end{pmatrix} \qquad \text{and} \qquad h=\begin{pmatrix}
 0 & 1\\
 -1 & 0
 \end{pmatrix},
\]
which contains $\ZZ/2n\ZZ$ as a subgroup of index two. The element $h$ defines an automorphism of~$\CC[u,v]^{Z/2n\ZZ}$ of order two and the invariant subalgebra is precisely the Kleinian singularity of type $D_{n - 1}$.
When the value of $n$ is not relevant,
we refer to such singularities as being of type~D
without reference to $n$.

Kleinian singularities are graded algebras. We can look at their noncommutative filtered deformations, that is, filtered associative algebras $\cA$
with associated graded algebra $\gr \cA$ equal to $A$,
and try to classify positive traces on these deformations.
Twisted and positive traces on deformations and $q$-deformations of Kleinian singularities of type $A$ were classified by the first author with Etingof, Rains, Stryker in~\cite{etingof2021twisted} and in~\cite{KQWeyl}. From a physical point of view, positive traces were studied in~\cite{ DFPY, DFPYCoulomb, DG,DPY}.

In the first part of the article, we consider the case of deformations of type $D$.
Not all deformations of Kleinian singularities of type $D$ can be obtained as a subalgebra inside a Kleinian singularity of type $A$, but in the first part we consider the ones that do. Moreover, we assume that the quantization parameter is generic.

We obtain classification results for traces
on deformations of Kleinian singularities of type D
analogous to those obtained in~\cite{etingof2021twisted} for type A.
Our most important result in this direction is the following theorem.

\begin{theorem*}[Theorem~\ref{thm:pos_trace}]
 Under assumptions described in Section~{\rm\ref{sec:pos}},
 all positive traces of filtered deformations of Kleinian singularities
 of type D
 are restrictions of positive traces
 of filtered deformations of Kleinian singularities of type A.
\end{theorem*}

We outline the technical preliminaries for our work on deformations of Kleinian singularities
in Section~\ref{sec:prelim}.
Then, in Section~\ref{sec:identities},
we introduce several algebra elements
which are important for our results
and prove various identities and relations among them.
We characterize traces of deformations
of Kleinian singularities of type~D
that satisfy the assumptions above in Section~\ref{sec:trace_classif},
and use this characterization
to obtain an analytic formula for traces
in Section~\ref{sec:analytic_form}.
Finally, in Section~\ref{sec:pos}, we prove our main result
regarding positive traces
of filtered deformations of Kleinian singularities of type D.

We now turn to the second part of the article.

Let $G$ be a complex reductive group and $N$ be a representation of $G$. One can construct a~supersymmetric gauge theory from this data. There are two algebraic varieties associated to this theory~-- the Higgs and Coulomb branches. A mathematical definition of Coulomb branches was given by Braverman, Finkelberg and Nakajima in~\cite{BFN}.

In the case when $N=0$, the Coulomb branch was computed earlier in~\cite{BFM}, it is isomorphic to the universal centralizer scheme \smash{$\mathfrak{Z}_{\check{\mathfrak{g}}}^{\check{G}}$} for the Langlands dual group. Other examples include slices in the affine Grassmannian for type $ADE$ simple Lie groups~\cite{BFNQuivers}, in particular the nilpotent cone~$\mc{N}_{\mathfrak{sl}_n}$ is a Coulomb branch. For the proof of the last assertion, see the last remark in~\cite{GW}, for example.

Each Coulomb branch $\mc{M}_{G,N}=\operatorname{Spec} \mc{A}_{G,N}$ comes with a quantization of its ring of functions~$\mc{A}_{G,N}^{\hbar}$. For example, any quantization of the nilpotent cone algebra $\CC[\mc{N}_{\mathfrak{sl}_n}]$ is a central reduction of $U(\mathfrak{sl}_n)$. This example provides a link between positive traces on quantized Coulomb branches and infinite-dimensional unitary representations of simple complex Lie groups. For an example of how this works in the case when $n=2$, see \cite[Section~4.3, pages~43 and~46]{klyuev2023unitarizability}.

In \cite[Lemma~6.9]{BFN}, it is proved that in the case of $G=\SL_2(\CC)$ Coulomb branches are exactly type $D_n$ Kleinian singularities $xy^2=z^2+x^{n-1}$ plus three exceptions: the cases $n=2$, $n=1$ of this equation, and the surface $xy^2=z^2+y$. If we take the representation $N=\bigl(\CC^2\bigr)^m$, we get~${n=m}$. In the notation of~\cite{BFN}, the representation is denoted by $\mathbf{N}$ and our $n$ corresponds to~$N$ in Lemma~6.9.
This fact motivates the title of the article, but we do not use any facts about Coulomb branches in the first part of the article.

The Coulomb branch algebra $\mc{A}_{G,N}$ is defined as equivariant homology of an infinite-dimen\-sion\-al scheme $\mc{R}_{G,N}$. Changing homology to K-theory, we also get an algebra, called the K-theoretic Coulomb branch algebra. Physically, we take 4-dimensional gauge theory with the same group~$G$ and matter $M$, and compactify it on a circle. Algebraically, a deformed $K$-theoretic Coulomb branch algebra $\mc{A}_{G,N}^q$ is a certain $q$-deformation of $\mc{A}_{G,N}$.

In the second part of the paper, we consider the algebra $\mc{A}$ introduced by Gaiotto and Teschner in \cite[Section~3.5]{GT}. This algebra contains $K$-theoretic Coulomb branches of pure~$\SL(2)$ and~$\PGL(2)$ gauge theories in the following sense. Gaiotto and Teschner use the description of Coulomb branch via the abelianization map, also known as the localization map in the mathematical literature. In our case, the target of the localization map is a certain localization $W$ of a $q$-Weyl algebra $W_0=\CC\langle u,v\rangle/(vw-qwv)$. Then $\mc{A}$ is generated by the images of \smash{$\mc{A}_{\SL(2),0}^q$} and~\smash{$\mc{A}_{\PGL(2),0}^q$} in $W$. In Section~\ref{SecAlgebra}, we will describe $\mc{A}$ and its connection to the BFN construction and show that there exist involutions $\tau_1$, $\tau_2$ of $\mc{A}$ such that \smash{$\mc{A}_{SL(2),0}^q=\mc{A}^{\tau_1}$} and \smash{$\mc{A}_{\PGL(2),0}^q=\mc{A}^{\tau_2}$}, the fixed point subalgebras.

While $\mc{A}$ does not exactly fit into the sequence of $q$-deformations of Kleinian singularities of type $D$, one could use a similar strategy to classify positive traces on deformed $K$-theoretic Coulomb branches with the gauge group $\SL(2)$ or $\PGL(2)$. Also, Kauffman skein algebras should be equal to the ``enlarged'' Coulomb branches constructed similarly to Section~\ref{SecAlgebra}. Allegretti and Shan proved in~\cite{AS} in the paragraph after Proposition~5.5 that the skein algebra for the once-punctured torus is isomorphic to \smash{$K^{\SL(2,\CC[t]\times \CC[t]^{\times})\rtimes \CC_q^{\times}}(\mc{R}_{\PGL(2),\mathfrak{sl}_2})$}. The latter algebra is the universal flavor deformation of a modified BFN construction in the case when $N=\mathfrak{sl}_2$ is the adjoint representation.

We classify positive traces on $\mc{A}$ with respect to an involution $\rho=\rho_m$ that depends on an integer $m$. In the case when $m=2$ this involution coincides with the one considered by Gaiotto and Teschner, and we further prove that the positive trace is unique up to scaling.

In Section~\ref{SecAlgebra}, we describe $\mc{A}$ and prove a proposition that describes the full image of $\mc{A}$ in the localized $q$-Weyl algebra $W$. In Section~\ref{SecTwistedTraces}, we describe traces in terms of a quasi-periodic function~$\omega$ holomorphic on $\CC^{\times}$. In Section~\ref{SecPositiveTraces}, we express the positivity condition on $T$ in terms of the behavior of $\omega$ on $S^1$ and $\sqrt{q}S^1$, finishing the classification. The answer is given by the following theorem.

\begin{theorem*}[Theorem~\ref{ThrAnswerKTheory}]
 Positive traces on $\mc{A}$ are in one-to-one correspondence with holomorphic functions $\omega$ on $\CC^{\times}$ such that
 \begin{enumerate}\itemsep=0pt
 \item[$(1)$] $\omega(qz)=q^{-\frac{m}{2}}z^{-m}\omega(z)$;
 \item[$(2)$] $\omega(z)=\omega\bigl(z^{-1}\bigr)$;
 \item[$(3)$] $\omega(1)=\omega(-1)=0$;
 \item[$(4)$] $\omega(z)$ and $z^{-\frac{m}{2}}\omega\bigl(\sqrt{q^{-1}}z\bigr)$ are nonnegative on $S^1$.
 \end{enumerate}

 The convex cone of such functions $\omega$ has real dimension $\frac{m}{2}-1$. In particular, for $m=4$ there is a unique positive trace up to a constant.
 \end{theorem*}

\pdfbookmark[1]{Part 1: Deformations of type D}{part1}

\begin{center}
 {\huge \bf Part 1: Deformations of type $\boldsymbol{D}$}
\end{center}

\section{Preliminaries}
\label{sec:prelim}

Throughout Sections~\ref{sec:prelim}--\ref{sec:pos},
we fix a positive even integer $n$ with $m = \frac{n}{2}$
and denote $\eps = \exp\bigl(\rn\bigr)$.
For a finite subgroup $G$
of the special linear group $\SL_2(\CC)$,
the Kleinian singularity $\CC^2 / G$
has as its coordinate ring
the commutative graded algebra $A = \CC[u, v]^G$
of $G$-invariant polynomials in two variables,
with the action of $G$ on $A$ defined by
\begin{equation*}
 \begin{bmatrix}
 a & b \\
 c & d
 \end{bmatrix}
 \cdot P(u, v) = P(au + bv, cu + dv).
\end{equation*}
We are interested in the case when $G = \DD_n$,
the dicyclic group generated by elements
\begin{equation*}
 g = \begin{bmatrix}
 \eps & 0 \\
 0 & \eps^{-1}
 \end{bmatrix}, \qquad
 h = \begin{bmatrix}
 0 & 1 \\
 -1 & 0
 \end{bmatrix}
\end{equation*}
with $g^n = h^4 = ghgh^{-1}$
all equal to the group identity
and $g^m = h^2$.
The group $\DD_n$ acts on~$\CC[u, v]$ by
$g \acts u = \eps u$, $g \acts v = \eps^{-1} v$,
$h \acts u = v$, $h \acts v = -u$,
and the invariant polynomials generating $A$
are $u^2 v^2$, $u^n + v^n$, and $uv(u^n - v^n)$.

All elements of the algebra $A$ are sums of homogeneous polynomials,
so $A$ is graded by the degree of homogeneous polynomials.
We consider filtered deformations
$\cA$ of $A$, that is, filtered associative algebras $\cA$
with associated graded algebra $\gr \cA$ equal to $A$.
In particular, we make use of the Crawley--Boevey--Holland~\cite{cbh}
filtered deformations.
Let $\CC[x, y] \# G$ be the skew group algebra
constructed from the vector space $\CC[x, y] \otimes \CC[G]$
with multiplication given by
\begin{equation*}
 (P \otimes g)(Q \otimes h) = Pg(Q) \otimes gh,
\end{equation*}
where $P$ and $Q$ are polynomials, $g$ and $h$ are group elements,
and $g(Q)$ denotes the action of $g$ on $Q$.
For an element $c$ of the center $Z(\CC[G])$ of the group algebra, let
\begin{equation*}
 H_c = (\CC[x, y] \# G)/(xy - yx - c).
\end{equation*}
For the idempotent \smash{$e = \frac1{|G|} \sum_{g \in G} g$},
let $\cO_c = eH_ce$, which is an algebra with unit $e$.
Losev~\cite[Theorem~3.4]{losev2022deformations}
proved that
for any finite subgroup $G$ of $\SL_2(\CC)$,
any filtered deformation of~$\CC[x, y]^G$
is isomorphic to some algebra $\cO_c$
with $c \in Z(\CC[G])$.

\begin{Definition}
 A trace on $\cA$ is a linear function $T \colon \cA \rightarrow \CC$
 satisfying $T(ab) = T(ba)$ for all~${a, b \in \cA}$.
\end{Definition}

We only consider traces on noncommutative algebras,
so by~\cite[Theorem~0.4\,(1)]{cbh},
we may assume the coefficient of $c$ on the group identity is nonzero.
Also, by~\cite[Theorem~0.4\,(2)]{cbh} that
for $c$ outside a finite set of hyperplanes, the algebra $\cO_c$ is Morita equivalent to $H_c$. We consider only such $c$, which we call \emph{generic}. We claim that for generic $c$ any trace on $\cO_c$ is a restriction of
a trace on $H_c$.

This implication of Theorem~0.4\,(2) is proved in Corollary~6.4, in particular the Morita equivalence is given by bimodules $eH_c$ and $H_ce$. Now, for any two Morita equivalent algebras $A$ and~$B$ with Morita equivalence given by an $(A-B)$-bimodule $M$ and a $(B-A)$-bimodule $N$ there is one-to-one correspondence between traces on $A$ and $B$ given by
\[
T_B\bigg(\sum_i n_i\otimes_A m_i\bigg)=T_A\bigg(\sum_i m_i\otimes_B n_i\bigg).
\]
 The map $T_B$ is well-defined because $T_A$ is a trace and the map $T_B$ is a trace because $T_A$ is well-defined. Applying this to $A=H_c$, $B=\cO_c$, we get
 \[
 T_B\bigg(\sum n_i\otimes_A m_i\bigg)=T_A\bigg(\sum m_i\otimes_B n_i\bigg)=T_A\bigg(\sum m_in_i\bigg)=T_A\bigg(\sum n_im_i\bigg),
 \]
 where we used that $T_A$ is a trace in the last step. Now, $\sum n_i\otimes_A m_i$ corresponds to just $\sum n_im_i$, hence $T_B$ is the restriction of $T_A$ to $\cO_c$.

For a classification of traces on $\cO_c$,
it therefore suffices to characterize traces on $H_c$.

Letting $C_n$ denote the cyclic group of $n$ elements,
we further assume that $c$ is in $\CC[C_n]$.
Then in the algebra $H_c$, the following hold:
\begin{gather*}
 gx = \eps xg; \qquad
 gy = \eps^{-1} yg; \qquad
 hx = yh; \\
 hy = -xh; \qquad
 xy - yx = c = \sum_{i = 0}^{n - 1} c_i g^i,\qquad
 \text{with $c_0 \neq 0$},\
 c_i = c_{-i} \text{ for all $i$}.
\end{gather*}
Here the equality $c_i = c_{-i}$ holds because
$c$ is in $Z(\CC[\DD_n])$
and the conjugacy classes of $\DD_n$ are of the form
$\big\{g^i, g^{-i}\big\}$, $\big\{ g^{2i} h \mid i \in \ZZ \big\}$,
and $\big\{ g^{2i + 1} h \mid i \in \ZZ \big\}$.

We close this section with a definition of positive traces.

\begin{Definition}
 For an algebra $B$ and an antilinear automorphism
 $\rho \colon B \to B$ such that $\rho^2 = \id$,
 a trace $T$ on $B$ is said to be \emph{positive} if
 $T(a \rho(a)) > 0$ for all nonzero $a$ in $B$.
\end{Definition}

The antilinear automorphism we study
is defined on $H_c$ by
$\rho(x) = y$, $\rho(y) = -x$, $\rho(g) = g$, $\rho(h) = h$.
We return to positive traces in Section~\ref{sec:pos}.

\section{Algebra identities and relations}
\label{sec:identities}

In this section, we introduce several important algebra elements
and describe relations between them.
The proofs are routine, but we provide them for the sake of completion.

Recall that $c = \sum_{i = 0}^{n - 1} c_i g^i$,
with $g$ a generator of $\DD_n$ such that $g^n$ is the group identity,
and~${\eps = \exp\bigl(\rn\bigr)}$.
Let
\begin{equation*}
 k = \frac1{c_0} \left( xy + \sum\limits_{i = 1}^{n - 1}
 \frac{c_i g^i}{\eps^i - 1} \right)
 - \half.
\end{equation*}
The following lemma describes commutativity relations
involving $k$
which are important for our results.

\begin{Lemma}
 \label{lem:k_prop}
 The following equalities hold in $H_c$:
 \begin{itemize}\itemsep=0pt
 \item $gk = kg$;
 \item $hk = -kh$;
 \item $[k, x] = -x$;
 \item $kx = x(k - 1)$;
 \item $[k, y] = y$;
 \item $ky = y(k + 1)$.
 \end{itemize}
\end{Lemma}

\begin{proof}
 Because $gxy = \eps xgy = xyg$,
 it is clear that $g$ and $k$ commute.
 Also, we have
 \begin{align*}
 hk ={}& \frac1{c_0} \left( hxy
 + \sum\limits_{i = 1}^{n - 1}
 \frac{h c_i g^i}{\eps^i - 1} \right) - \frac{h}{2}= \frac1{c_0} \left( -yx h
 + \sum\limits_{i = 1}^{n - 1}
 \frac{c_i g^{-i} h}{\eps^i - 1} \right) - \frac{h}{2} \\
 ={}& \frac1{c_0} \left( c - xy
 + \sum\limits_{i = 1}^{n - 1}
 \frac{c_i g^i}{\eps^{-i} - 1} \right) h - \frac{h}{2}= - \frac1{c_0} \left( xy
 + \sum\limits_{i = 1}^{n - 1}
 \frac{c_i g^i}{\eps^i - 1} \right) h + \frac{h}{2}= -kh.
 \end{align*}
 We now prove the commutation relations with $x$ and $y$.
 We have $xyx = x(xy - c) = x^2 y - xc$,
 so
 \begin{align*}
 [k, x]
 ={}& \frac1{c_0} \left( xyx - x^2 y
 + \sum\limits_{i = 1}^{n - 1}
 \frac{c_i g^i x - x c_i g^i}{\eps^i - 1} \right)= \frac1{c_0} \left( -xc
 + \sum\limits_{i = 1}^{n - 1}
 \frac{\eps^i x c_i g^i - x c_i g^i}{\eps^i - 1} \right) \\
 ={}& \frac1{c_0} \left( -xc
 + x \sum\limits_{i = 1}^{n - 1} c_i g^i \right) = -x.
 \end{align*}
 It follows immediately that $kx = x(k - 1)$,
 and the commutation relations with $y$
 follow by conjugating the commutation relations with $x$
 by $h$.
\end{proof}

Now let
\begin{equation*}
 e_q = \frac1n \sum\limits_{i = 0}^{n - 1} \eps^{iq} g^i
\end{equation*}
for integers $q$.
Note that $e_q = e_{q + n}$ for all $q$.
The elements $e_q$ form an important basis for $\CC[C_n]$,
and it is often convenient to express elements of $H_c$
using $e_q$.
The following lemma describes relations
involving $e_q$.

\begin{Lemma}
 \label{lem:e_prop}
 The following equalities hold in $H_c$:
 \begin{itemize}\itemsep=0pt
 \item $e_q g^i = g^i e_q = \eps^{-iq} e_q$;
 \item $h e_q = e_{-q} h$;
 \item $e_q x = x e_{q + 1}$;
 \item $e_q y = y e_{q - 1}$;
 \item $e_q k = k e_q$;
 \item $e_p e_q = \delta_{p, q} e_q$,
 where $\delta$ is the Kronecker delta.
 \end{itemize}
\end{Lemma}

\begin{proof}
 Clearly, $g^i$ and $e_q$ commute.
 We have
 \begin{equation*}
 g^i e_q = \frac1n \sum\limits_{j = 0}^{n - 1}
 \eps^{jq} g^{i + j}
 = \frac1n \sum\limits_{j = 0}^{n - 1}
 \eps^{(j - i)q} g^j
 = \eps^{-iq} e_q
 \end{equation*}
 and
 \begin{equation*}
 h e_q = \frac1n \sum\limits_{i = 0}^{n - 1}
 \eps^{iq} h g^i
 = \frac1n \sum\limits_{i = 0}^{n - 1}
 \eps^{iq} g^{-i} h
 = e_{-q} h,
 \end{equation*}
 proving the commutation relations with group elements.
 Also,
 \begin{equation*}
 e_q x = \frac1n \sum\limits_{i = 0}^{n - 1}
 \eps^{iq} g^i x
 = \frac1n \sum\limits_{i = 0}^{n - 1}
 \eps^{iq + i} x g^i
 = x e_{q + 1},
 \end{equation*}
 and conjugating by $h$,
 we obtain $e_q y = y e_{q - 1}$.
 It follows from the established commutation relations
 that $e_q$ commutes with $xy$ and each $g^i$,
 so $e_q$ commutes with $k$.
 Finally, we have
 \begin{equation*}
 e_p e_q = \frac1n \sum\limits_{i = 0}^{n - 1}
 \eps^{ip} g^i e_q
 = \frac1n \sum\limits_{i = 0}^{n - 1}
 \eps^{i(p - q)} e_q.
 \end{equation*}
 If $p = q$, then all terms in the sum are equal to $e_q$,
 so $e_p e_q = e_q$.
 If $p \neq q$, then $\sum_{i = 0}^{n - 1} \eps^{i(p - q)} = 0$,
 so $e_p e_q = 0$.
\end{proof}

Finally, we define
\begin{equation*}
 \alpha_q = \frac1{c_0}
 \sum\limits_{i = 1}^{n - 1} \frac{c_i \eps^{-iq}}{\eps^i - 1}
\end{equation*}
for integers $q$,
and once again observe that $\alpha_{q} = \alpha_{q + n}$ for all $q$.
The elements $\alpha_q$ have an important symmetry
and allow for useful transitions
from complicated expressions involving $e_q$
to scalar multiples of $e_q$,
as shown by the following lemma.

\begin{Lemma}
 \label{lem:a_prop}
 The following statements hold in $H_c$:
 \begin{itemize}\itemsep=0pt
 \item $\alpha_q = -\alpha_{-q - 1}$;
 \item the elements $\alpha_p$ and $e_q$ satisfy
 \begin{equation*}
 \alpha_p e_q = \frac1{c_0}
 \left( \sum_{i = 1}^{n - 1}
 \frac{c_i \eps^{i(q - p)} g^i}{\eps^i - 1}
 \right) e_q.
 \end{equation*}
 \end{itemize}
\end{Lemma}

\begin{proof}
 We have
 \begin{align*}
 \alpha_q ={}& \frac1{c_0}
 \sum\limits_{i = 1}^{n - 1}
 \frac{c_i \eps^{-iq}}{\eps^i - 1}
 = \frac1{c_0} \sum\limits_{i = 1}^{n - 1}
 \frac{c_{-i} \eps^{-iq}}{\eps^i - 1}
 = \frac1{c_0} \sum\limits_{i = 1}^{n - 1}
 \frac{c_i \eps^{iq}}{\eps^{-i} - 1}
 = \frac1{c_0} \sum\limits_{i = 1}^{n - 1}
 \frac{c_i \eps^{i (q + 1)}}{1 - \eps^i}\\
 ={}& -\alpha_{-q - 1}.
 \end{align*}
 We also have
 \begin{align*}
 \alpha_p e_q
 ={}& \frac1{c_0} \sum\limits_{i = 1}^{n - 1}
 \frac{c_i \eps^{-ip} e_q}{\eps^i - 1} = \frac1{c_0} \sum\limits_{i = 1}^{n - 1}
 \frac{c_i \eps^{i(q - p)}
 \eps^{-iq} e_q}{\eps^i - 1} = \frac1{c_0} \left( \sum\limits_{i = 1}^{n - 1}
 \frac{c_i \eps^{i (q - p)} g^i}{\eps^i - 1}
 \right) e_q,
 \end{align*}
 proving the lemma.
\end{proof}

The elements $e_q$ are a basis for $\CC[G]$,
so we may write
\begin{equation*}
 k = \frac1{c_0} \Bigg( xy -
 \sum\limits_{q = 0}^{n - 1} \beta_q e_q \Bigg),
 \qquad \textrm{or} \qquad
 xy = c_0 k + \sum\limits_{q = 0}^{n - 1} \beta_q e_q,
\end{equation*}
for complex numbers $\beta_q$
and extend the $\beta_q$ modulo $n$,
so that $\beta_q = \beta_{q + n}$ for all integers $q$.
More generally, we have the following lemma.

\begin{Lemma}
 \label{lem:xy_poly}
 For all positive integers $a$,
 the elements $x^a y^a$ and $y^a x^a$
 are polynomials in $k$ and group elements
 given by
 \begin{align*}
 x^a y^a ={}& \prod_{i = 0}^{a - 1} \Bigg( c_0 (k + i)
 + \sum\limits_{q = 0}^{n - 1} \beta_{q + i} e_q \Bigg),
 \\
 y^a x^a ={}& (-1)^a \prod\limits_{i = 0}^{a - 1}
 \Bigg( c_0 (-k + i)
 + \sum\limits_{q = 0}^{n - 1} \beta_{q + i} e_{-q}
 \Bigg).
 \end{align*}
 Also, the elements $x^a y^a e_q$ and $y^a x^a e_q$
 can be written as polynomial expressions in $k$
 multiplied by $e_q$ as follows
 \begin{align*}
 x^a y^a e_q ={}& e_q \prod\limits_{i = 0}^{a - 1}
 (c_0 (k + i) + \beta_{q + i}), \\
 y^a x^a e_q ={}& (-1)^a e_q \prod\limits_{i = 0}^{a - 1}
 (c_0 (-k + i) + \beta_{-q + i}).
 \end{align*}
\end{Lemma}

\begin{proof}
 We prove the polynomial expression for $x^a y^a$
 by induction,
 with the base case of $a = 1$ true by definition.
 Assume that
 $x^a y^a e_q = e_q \prod_{i = 0}^{a - 1}
 \bigl(c_0 (k + i) + \sum_{q = 0}^{n - 1} \beta_{q + i} e_q\bigr)$.
 Then
 \begin{align*}
 x^{a + 1} y^{a + 1}
 ={}& x^a \Bigg( c_0 k
 + \sum\limits_{q = 0}^{n - 1}
 \beta_q e_q \Bigg) y^a = x^a y^a \Bigg( c_0 (k + a)
 + \sum\limits_{q = 0}^{n - 1}
 \beta_q e_{q - a} \Bigg) \\
 ={}& x^a y^a \Bigg( c_0 (k + a)
 + \sum\limits_{q = 0}^{n - 1}
 \beta_{q + a} e_q \Bigg) = \prod\limits_{i = 0}^a
 \Bigg( c_0 (k + i)
 + \sum\limits_{q = 0}^{n - 1}
 \beta_{q + i} e_q \Bigg),
 \end{align*}
 as needed.
 The analogous claim for $y^a x^a$
 follows from conjugation by $h$.
 The expressions for~$x^a y^a e_q$ and $y^a x^a e_q$
 follow using the statement that $e_p e_q = \delta_{p, q} e_q$
 from Lemma~\ref{lem:e_prop}.
\end{proof}

The numbers $\beta_q$ are related to the $\alpha_q$
by the following lemma.

\begin{Lemma}
 \label{lem:a_b_rel}
 For all $q$, we have $\beta_q = \frac{c_0}{2} - c_0 \alpha_q$.
\end{Lemma}

\begin{proof}
 By Lemma~\ref{lem:a_prop}, we have
 \begin{align*}
 xy e_q ={}& \Bigg( c_0 k + \frac{c_0}{2}
 - \sum\limits_{i = 1}^{n - 1}
 \frac{c_i g^i}{\eps^i - 1} \Bigg) e_q = \left( c_0k + \frac{c_0}{2} - c_0 \alpha_q \right) e_q.
 \end{align*}
 Also, by definition,
 $xy e_q = ( c_0 k + \beta_q ) e_q$.
 Thus
 \begin{align*}
 \left( c_0 k + \beta_q \right) e_q
 = \left( c_0 k + \frac{c_0}{2}
 - c_0 \alpha_q \right) e_q ,\qquad
 \beta_q e_q
 = \left( \frac{c_0}{2}
 - c_0 \alpha_q \right) e_q,
 \end{align*}
 whence it follows that $\beta_q = \frac{c_0}{2} - c_0 \alpha_q$.
\end{proof}

\section{Characterization of traces}
\label{sec:trace_classif}

In this section, we study traces using basis elements of the form
$x^i R(k) e_q$, $y^i R(k) e_q$, $x^i R(k) e_q h$, and $y^i R(k) e_q h$,
with $R(k) \in \CC[k]$.
Let $T$ be a trace on $H_c$.
We begin with a lemma about the values $T(e_q h)$.

\begin{Lemma}
 \label{lem:trace_eh_zero}
 If $m \nmid q$, then $T(e_q h) = 0$.
\end{Lemma}

\begin{proof}
 It follows from the definition of a trace that
 the value of $T$ must be equal
 on all conjugacy classes of $\DD_n$.
 This gives
 \begin{align*}
 T(e_q h) ={}& \frac1n T\left(
 \sum\limits_{i = 0}^{n - 1}
 \eps^{iq} g^i h \right) = \frac1n \left( \sum\limits_{i = 0}^{m - 1}
 \eps^{2iq} \right) T(h)
 + \frac1n \left( \sum\limits_{i = 0}^{m - 1}
 \eps^{(2i + 1)q} \right) T(gh) \\
 ={}& \frac1n \left( \sum\limits_{i = 0}^{m - 1}
 \eps^{2iq} \right) T(h)
 + \frac1n \eps^q \left( \sum\limits_{i = 0}^{m - 1}
 \eps^{2iq} \right) T(gh).
 \end{align*}
 If $m \nmid q$, then $n \nmid 2q$, so the sums of roots of unity
 are equal to $0$.
 Therefore, $T(e_q h)$ is nonzero only if $m \mid q$.
\end{proof}

The following theorem classifies traces on $H_c$.
In particular,
conditions~\ref{cond:inv}--\ref{cond:k_poly}
express trace values on $H_c$
in terms of trace values on elements of the form $R(k) e_q$,
with $R$ a polynomial,
and conditions~\ref{cond:shift} and~\ref{cond:even}
restrict possible trace values on such terms.

\begin{Theorem}
 \label{thm:classif}
 A linear map $T\colon H_c \to \CC$
 is a trace on $H_c$ if and only if
 all of the following conditions hold:
 \begin{enumerate}[label=\numbers]\itemsep=0pt
 \item $T$ is invariant under conjugation
 by elements of $\DD_n$;
 \label{cond:inv}
 \item $T\bigl(x^i R(k) e_q\bigr) = T\bigl(y^i R(k) e_q\bigr) = 0$
 when $i \neq 0$;
 \label{cond:g0}
 \item $T\bigl(x^i R(k) e_q h\bigr) = T\bigl(y^i R(k) e_q h\bigr) = 0$
 when $2 \nmid i$;
 \label{cond:h0}
 \item $T\bigl(x^{2i} R(k) e_q h\bigr) = T(S(k) R(k + i) e_{q - i} h)$,
 where $S$ is the polynomial
 satisfying that $x^i y^i e_{q - i}\allowbreak = S(k) e_{q - i}$;
 \label{cond:h_av}
 \item $T\bigl(y^{2i} R(k) e_q h\bigr) = (-1)^i T(S(k) R(k - i) e_{q + i} h)$,
 where $S$ is the polynomial
 satisfying that $y^i x^i e_{q + i} = S(k) e_{q + i}$;
 \label{cond:h_av_flip}
 \item $T\bigl(k^i e_q h\bigr) = 0$ for $i > 0$;
 \label{cond:k_poly}
 \item $T\bigl( \bigl(k + \half - \alpha_q\bigr) R\bigl(k + \half\bigr) e_q\bigr) =
 T\bigl( \bigl(k - \half - \alpha_q\bigr) R\bigl(k - \half\bigr) e_{q + 1}\bigr)$
 for all $q \in \ZZ$ and $R \in \CC[X]$;
 \label{cond:shift}
 \item $T( R(k) e_q ) = T( R(-k) e_{-q} )$
 for all $q \in \ZZ$ and $R \in \CC[X]$.
 \label{cond:even}
 \end{enumerate}
\end{Theorem}

\begin{proof}
 Assume first that $T$ is a trace;
 we will show that the listed conditions must hold.
 \begin{enumerate}[label=\numbers]\itemsep=0pt
 \item This is part of the definition of a trace.
 \item The trace condition gives
 $0 = T\bigl(\big[k, x^i R(k) e_q\big]\bigr) = -i T\bigl(x^i R(k) e_q\bigr)$,
 and similarly
 $0 = T\bigl(\big[k, y^i R(k) e_q\big]\bigr) = i T\bigl(y^i R(k) e_q\bigr)$,
 from which the claim follows.
 \item Conjugation by $h^2$ multiplies both $x$ and $y$ by $-1$
 while fixing $e_q$ and $h$, so
 \begin{gather*}
 T\bigl(x^i R(k) e_q h\bigr) = (-1)^i T\bigl(x^i R(k) e_q h\bigr) \qquad \text{and}\\
T\bigl(y^i R(k) e_q h\bigr) = (-1)^i T\bigl(y^i R(k) e_q h\bigr),\end{gather*}
 proving the claim.
 \item Using the trace condition
 to commute $x^i$ and $x^i R(k) e_q h$,
 we obtain
 \begin{align*}
 T\bigl(x^{2i} R(k) e_q h\bigr)
 ={}& T\bigl(x^i R(k) e_q h x^i\bigr)= T\bigl(x^i R(k) e_q y^i h\bigr) \\
 ={}& T\bigl(x^i y^i R(k + i) e_{q - i} h\bigr) = T\bigl(x^i y^i e_{q - i} R(k + i) h\bigr) \\
 ={}& T(S(k) e_{q - i} R(k + i) h) = T(S(k) R(k + i) e_{q - i} h),
 \end{align*}
 as desired.
 \item Similarly to the proof of condition~\ref{cond:h_av},
 we obtain
 \begin{align*}
 T\bigl(y^{2i} R(k) e_q h\bigr)
 ={}& T\bigl(y^i R(k) e_q h y^i\bigr)= (-1)^i T\bigl(y^i R(k) e_q x^i h\bigr) \\
 ={}& (-1)^i T\bigl(y^i x^i R(k - i) e_{q + i} h\bigr) = (-1)^i T(S(k) R(k - i) e_{q + i} h),
 \end{align*}
 as needed.
 \item The trace condition gives
 $0 = T\bigl(\big[k, k^{i - 1} e_q h\big]\bigr) = T\bigl(2 k^i e_q h\bigr)$,
 from which the claim follows.
 \item Using Lemma~\ref{lem:a_prop}
 and the commutativity of $e_q$ and $k$, we obtain
 \begin{align*}
 T\bigl( \bigl( k + \thalf - \alpha_q \bigr)
 R\bigl( k + \thalf \bigr) e_q \bigr)
 ={}& T\left( \left( k + \half -
 \frac1{c_0} \sum\limits_{i = 1}^{n - 1}
 \frac{c_i g^i}{\eps^i - 1} \right)
 R\bigl( k + \thalf \bigr) e_q \right) \\
 ={}& \frac1{c_0}
 T\bigl( xy R\bigl( k + \thalf \bigr) e_q \bigr).
 \end{align*}
 Using the trace condition
 to commute $x$ and $y R\bigl(k + \half\bigr) e_q$,
 we find
 \begin{align*}
 T\bigl( \bigl( k + \thalf - \alpha_q \bigr)
 R\bigl( k + \thalf \bigr) e_q \bigr)
 ={}& \frac1{c_0}
 T\bigl( y R\bigl( k + \thalf \bigr) e_q x \bigr)
 = \frac1{c_0}
 T\bigl( yx R\bigl( k - \thalf \bigr)
 e_{q + 1} \bigr) \\
 ={}& T\left( \left( k - \half -
 \frac1{c_0} \sum\limits_{i = 1}^{n - 1}
 \frac{c_i \eps^i g^i}{\eps^i - 1} \right)
 R\bigl( k - \thalf \bigr) e_{q + 1} \right),
 \end{align*}
 and using Lemma~\ref{lem:a_prop}
 and the commutativity of $e_q$ and $k$
 once again, we obtain
 \begin{align*}
 T\bigl( \bigl( k + \thalf - \alpha_q \bigr)
 R\bigl( k + \thalf \bigr) e_q \bigr)
 ={}& T\bigl( \bigl( k - \thalf - \alpha_q \bigr)
 R\bigl( k - \thalf \bigr) e_{q + 1} \bigr),
 \end{align*}
 as desired.
 \item This follows from conjugation by $h$.
 \end{enumerate}

 We now prove the converse.
 Suppose $T$ is a linear map from $H_c$ to $\CC$
 satisfying the listed conditions.
 We wish to show that for all $a$ in $H_c$,
 we have
 \[
 T([x, a]) = T([y, a]) = T([g, a]) = T([h, a]) = 0.
 \]
 The equality $T([g, a]) = T([h, a]) = 0$ follows
 from condition~\ref{cond:inv},
 so it remains to verify the result for commutation by $x$ and $y$.
 Also, the statements that $T([y, a]) = 0$ for all $a$
 follows from the statement that $T([x, a]) = 0$ for all $a$
 by conjugation by $h$ and condition~\ref{cond:inv}.
 It therefore suffices to show that $T([x, a]) = 0$
 when $a$ is equal to a basis element
 of the form~$x^i R(k) e_q$, $y^i R(k) e_q$,
 $x^i R(k) e_q h$, or $y^i R(k) e_q h$.
 We now check all possible cases.
 \begin{itemize}\itemsep=0pt
 \item Let $a = x^i R(k) e_q$ with $i$ nonnegative
 or $a = y^i R(k) e_q$ with $i > 1$.
 Then $T(xa) = 0 = T(ax)$ by
 condition~\ref{cond:g0}.
 \item Let $a = y R(k) e_q$.
 Then, using Lemma~\ref{lem:a_prop}, we obtain
 \begin{align*}
 T(xa) ={}& T\left( xy R(k) e_q \right) = c_0 T\left( \left( k + \half -
 \frac1{c_0} \sum\limits_{i = 1}^{n - 1}
 \frac{c_i g^i}{\eps^i - 1} \right)
 R(k) e_q \right) \\
 ={}& c_0 T\bigl( \bigl( k + \thalf
 - \alpha_q \bigr) R(k) e_q \bigr).
 \end{align*}
 Using condition~\ref{cond:shift}
 and Lemma~\ref{lem:a_prop},
 we obtain
 \begin{align*}
 T(xa) ={}& c_0
 T\bigl( \bigl( k - \thalf - \alpha_q \bigr)
 R(k - 1) e_{q + 1} \bigr) \\
 ={}& c_0 T\left( \left( k - \half -
 \frac1{c_0} \sum\limits_{i = 1}^{n - 1}
 \frac{c_i \eps^i g^i}{\eps^i - 1} \right)
 R(k - 1) e_{q + 1} \right) \\
 ={}& T\left( yx R(k - 1) e_{q + 1} \right)= T\left( y R(k) e_q x \right)= T(ax),
 \end{align*}
 as desired.
 \item Let $a = x^i R(k) e_q h$
 or $a = y^i R(k) e_q h$,
 with $i$ even.
 Then by condition~\ref{cond:h0},
 $T(xa) = 0 = T(ax)$.
 \item Let $a = x^{2i - 1} R(k) e_q h$,
 and let $S(X)$ be the polynomial
 such that $S(k) e_{q - 1} = xy e_{q - 1}$.
 Then by condition~\ref{cond:h_av},
 \begin{align*}
 T(xa) ={}& T\bigl( x^{2i} R(k) e_q h \bigr)= T\bigl( x^i y^i R(k + i)
 e_{q - i} h \bigr) \\
 ={}& T\bigl( x^{i - 1} (xy) y^{i - 1} R(k + i)
 e_{q - i} h \bigr) = T\bigl( x^{i - 1} (xy) y^{i - 1}
 e_{q - i} R(k + i) h \bigr) \\
 ={}& T\bigl( x^{i - 1} (xy) e_{q - 1} y^{i - 1}
 R(k + i) h\bigr) = T\bigl( x^{i - 1} S(k) e_{q - 1} y^{i - 1}
 R(k + i) h \bigr) \\
 ={}& T\bigl( x^{i - 1} y^{i - 1}
 S(k + i - 1) R(k + i) e_{q - i} h\bigr) = T\bigl( x^{2i - 2} S(k)
 R(k + 1) e_{q - 1} h \bigr) \\
 ={}& T\bigl( x^{2i - 2} (xy)
 R(k + 1) e_{q - 1} h \bigr) = T\bigl( x^{2i - 1}
 R(k) e_q y h \bigr) \\
 ={}& T\bigl(x^{2i - 1} R(k) e_q h x\bigr) = T(ax),
 \end{align*}
 as needed.
 \item Let $a = y^{2i + 1} R(k) e_q h$,
 and let $S(X)$ be the polynomial
 such that $S(k) e_{q + 2i} = xy e_{q + 2i}$.
 Then by condition~\ref{cond:h_av_flip},
 \begin{align*}
 T(xa) ={}& T\bigl( x y^{2i + 1}
 R(k) e_q h \bigr) = T\bigl( xy e_{q + 2i} y^{2i}
 R(k) h \bigr) \\
 ={}& T\bigl( S(k) e_{q + 2i} y^{2i}
 R(k) h \bigr) = T\bigl( y^{2i}
 S(k + 2i) R(k) e_q h \bigr) \\
 ={}& (-1)^i T\bigl( y^i x^i
 S(k + i) R(k - i) e_{q + i} h \bigr).
 \end{align*}
 On the other hand,
 \begin{align}
 T(ax) ={}& T\bigl( y^{2i + 1}
 R(k) e_q y h \bigr) = T\bigl( y^{2i + 2}
 R(k + 1) e_{q - 1} h \bigr) \nonumber\\
 ={}& (-1)^{i + 1} T\bigl( y^{i + 1} x^{i + 1}
 R(k - i) e_{q + i} h \bigr).\label{eq:t_ax}
 \end{align}
 So by condition~\ref{cond:k_poly},
 both $T(xa)$ and $T(ax)$
 are proportional to $T(e_{q + i} h)$,
 which, by Lemma~\ref{lem:trace_eh_zero},
 is nonzero only if $m \mid q + i$.
 We may therefore assume that $m \mid q + i$,
 because otherwise, $T(xa) = 0 = T(ax)$.
 It follows that $n \mid 2q + 2i$,
 so $\beta_{q + 2i} = \beta_{-q}$.
 Therefore, by Lemma~\ref{lem:xy_poly},
 \begin{align*}
 T(xa)
 ={}& (-1)^i T\bigl( y^i x^i
 S(k + i) R(k - i) e_{q + i} h \bigr) \\\
 ={}& T\Bigg( \Bigg(
 \prod\limits_{j = 0}^{i - 1}
 c_0 (-k + j) + \beta_{-q - i + j}
 \Bigg)
( c_0 (k + i) + \beta_{q + 2i})
 R(k - i) e_{q + i} h \Bigg) \\
 ={}& T\Bigg( \Bigg(
 \prod\limits_{j = 0}^{i - 1}
 c_0 (-k + j) + \beta_{-q - i + j}
 \Bigg)
( c_0 (k + i) + \beta_{-q})
 R(k - i) e_{q + i} h \Bigg).
 \end{align*}
 By condition~\ref{cond:k_poly},
 this expression depends only on its value
 for $k = 0$.
 Therefore, simplifying the expression for $T(xa)$
 and using \eqref{eq:t_ax}, we obtain
 \begin{align*}
 T(xa)
 ={}& T\Bigg( \Bigg(
 \prod\limits_{j = 0}^{i - 1}
 c_0 (-k + j) + \beta_{-q - i + j}
 \Bigg)
( c_0 (-k + i) + \beta_{-q})
 R(k - i) e_{q + i} h \Bigg) \\
 ={}& T\Bigg( \Bigg(
 \prod\limits_{j = 0}^i
 c_0 (-k + j) + \beta_{-q - i + j}
 \Bigg)
 R(k - i) e_{q + i} h \Bigg) \\
 ={}& (-1)^{i + 1} T\bigl( y^{i + 1} x^{i + 1}
 R(k - i) e_{q + i} h \bigr) = T(ax),
 \end{align*}
 as needed.
 \end{itemize}
 This resolves all cases, completing the proof.
\end{proof}

Theorem~\ref{thm:classif} reduces traces on $H_c$ to traces on a subalgebra
and two additional parameters, as shown by the following corollary.

\begin{Corollary}
 \label{cor:unique}
 A trace is uniquely determined by its values
 on elements of the form $R(k) e_q$, with $R$ a polynomial,
 and on the conjugacy classes $g^{2i} h$ and $g^{2i + 1} h$
 of $\DD_n$.
 Also, any choice of linear function $T$ on the subalgebra of $H_c$
 spanned by elements of the form $R(k) e_q$
 which satisfies conditions~\ref{cond:shift} and~\ref{cond:even}
 of Theorem~{\rm\ref{thm:classif}}
 and of values $T\bigl(g^{2i} h\bigr)$ and $T\bigl(g^{2i + 1} h\bigr)$
 defines a unique trace on $H_c$
 through conditions~\ref{cond:inv}--\ref{cond:k_poly}
 of Theorem~{\rm\ref{thm:classif}}.
\end{Corollary}

\begin{proof}
 By Theorem~\ref{thm:classif}, a trace is uniquely determined
 by its values on terms $R(k) e_q$
 and on the conjugacy classes $g^{2i} h$ and $g^{2i + 1} h$.
 We show that any choice of linear function on terms $R(k) e_q$
 which satisfies conditions~\ref{cond:shift} and~\ref{cond:even}
 of Theorem~\ref{thm:classif}
 and of values $T\bigl(g^{2i} h\bigr)$ and $T\bigl(g^{2i + 1} h\bigr)$
 gives a valid trace.
 Indeed, values of $T$ on basis elements
 are determined by conditions~\ref{cond:g0}--\ref{cond:k_poly}
 of Theorem~\ref{thm:classif},
 so all that is left to check
 is that condition~\ref{cond:inv} of Theorem~\ref{thm:classif}
 does not lead to a contradiction.
 Using the fact that $ghg^{-1} = g^2 h$
 and $g^i e_q = \eps^{-iq} e_q$,
 we see that conjugation by~$g$ preserves
 the relations among basis elements
 given in conditions~\ref{cond:g0}--\ref{cond:k_poly}
 of Theorem~\ref{thm:classif}
 and preserves conjugacy classes of $\DD_n$.
 Conjugation by $h$ also preserves
 relationship among basis elements,
 so the trace is indeed well-defined.
\end{proof}

We now compute the dimension of the space of traces on $H_c$.

\begin{Proposition}
 \label{prop:dim}
 The dimension of the space of traces on $H_c$
 is $\frac{n}{2} + 2$.
\end{Proposition}

\begin{proof}
 By Corollary~\ref{cor:unique},
 it suffices to show that the dimension of the space
 of linear maps satisfying
 conditions~\ref{cond:shift} and~\ref{cond:even}
 of Theorem~\ref{thm:classif}
 is $\frac{n}{2}$.
 Let $V = \CC[X]^{\oplus n}$,
 and define the linear map $\varphi \colon V \to V$ by
 \begin{gather*}
 \begin{split}
& \varphi(R_0 (X), \dots, R_{n - 1} (X)) \\
& \qquad= \bigl(R_0 \bigl(X + \thalf\bigr) - R_{n - 1} \bigl(X - \thalf\bigr), \dots,
 R_{n - 1} \bigl(X + \thalf\bigr) - R_{n - 2} \bigl(X - \thalf\bigr)\bigr),
\end{split}
 \end{gather*}
 where the $i$th component is equal to
 $R_i \bigl(X + \half\bigr) - R_{i - 1} \bigl(X - \thalf\bigr)$.
 Define the ideal $I$ of $V$ by
 \begin{equation*}
 I = \{ ((X - \alpha_0) R_0(X), \dots,
 (X - \alpha_{n - 1}) R_{n - 1}(X)) \mid R_i(X) \in \CC[X] \}
 \end{equation*}
 and the subspace $W$ of $V$ as the set of elements
 $(R_0(X), \dots, R_{n - 1}(X))$
 for which $R_i(X) = -R_{-i}(-X)$.
 The space of linear maps satisfying
 conditions~\ref{cond:shift} and~\ref{cond:even}
 of Theorem~\ref{thm:classif}
 is isomorphic to the vector space
 $(V / \langle \varphi(I), W \rangle)^*$,
 where $\langle \varphi(I), W) \rangle$
 denotes the linear span of~$\varphi(I)$ and $W$.
 We therefore wish to show that the dimension of
 $V / \langle \varphi(I), W \rangle$ is $\frac{n}{2}$.

 First, note that the map $\varphi$ is surjective.
 Indeed, to solve the equation
 $\varphi((R_0, \dots, R_{n - 1})) = (P_0, \dots, P_{n - 1})$,
 let $R_{n - 1}$ be a solution to the equation
 \[R_{n - 1} \bigl(X + n - \thalf\bigr) - R_{n - 1} \bigl(X - \thalf\bigr)
 = \sum_{i = 0}^{n - 1} P_i(X + i).\]
 Then $R_0$, \dots, $R_{n - 2}$ are uniquely defined
 by the first $n - 1$ components of the equation,
 and the resulting solution is valid because
 the sum $\sum_{i = 0}^{n - 1} \bigl(R_i \bigl(X + i + \half\bigr)
 - R_{i - 1} \bigl(X + i - \half\bigr)\bigr)$ is equal to~${R_{n - 1} \bigl(X + n - \half\bigr) - R_{n - 1} \bigl(X - \half\bigr)}$,
 which equals $\sum_{i = 0}^{n - 1} P_i(X + i)$, as needed.
 We also note that the kernel of $\varphi$
 is $\{(z, \dots, z) \mid z \in \CC\}$.

 Letting
$
 U = \{ (R_0 (X), \dots, R_{n - 1} (X))
 \mid R_i (X) = R_{-i - 1} (-X) \}$,
 we claim that the preimage of~$W$ under $\varphi$ is $U$.
 Indeed, $\varphi(U) \subseteq W$
 because for any element $(R_0 (X), \dots, R_{n - 1} (X)) \in U$,
 the $i$th component of $\varphi(R_0 (X), \dots, R_{n - 1} (X))$
 is \[R_i \bigl(X + \thalf\bigr) - R_{i - 1} \bigl(X - \thalf\bigr)
 = R_{-i - 1} \bigl(-X - \thalf\bigr) - R_{-i} \bigl(-X + \thalf\bigr),\]
 which is equal to the negative of the $(n - i)$th component.
 Conversely, for any
 element $(R_0 (X), \allowbreak\dots, R_{n - 1} (X))$
 mapped to $W$ by $\varphi$,
 the element of $V$ with $i$th component equal to
 \[
 \frac{R_i (X) + R_{-i - 1} (-X)}{2}
 \] belongs to $U$
 and is also mapped to $W$ by $\varphi$.
 Therefore,
$V / \langle \varphi(I), W \rangle
 \cong V / \langle I, U \rangle$.

 Let $\psi \colon V \to V / I \cong \CC^n$ be the quotient map given by
 mapping $(R_0(X), \dots, R_{n - 1}(X))$
 to~$(R_0(\alpha_0), \dots, R_{n - 1}(\alpha_{n - 1}))$.
 Because $\alpha_i = -\alpha_{-i - 1}$,
 $\psi(U)$ is the set $S$ of vectors $(v_0, \dots, v_{n - 1})$
 such that $v_i = v_{-i - 1}$ for all $i$.
 Therefore,
$ V / \langle I, U \rangle \cong \CC^n / S$,
 the dimension of which is $\frac{n}{2}$,
 as desired.
\end{proof}

\section{Analytic formula for traces}
\label{sec:analytic_form}

We now give an analytic description of traces
using the notion of a good contour from~\cite{klyuev2023unitarizability}.
In particular, similar to \cite[Definition~3.6]{klyuev2023unitarizability},
we say that a non-self-intersecting curve $C$
with exactly one point in the set $\RR + ri$ for all real $r$
is a good contour if the following three conditions hold:%
\begin{enumerate}\itemsep=0pt
 \item[$(1)$] There exist real $r > 0$ and $a$ such that,
 letting $B_r$ be the disk of radius $r$ centered at the origin,
 $C \setminus B_r$ coincides with
 $(a + {\rm i}\RR_{>0}) \cup (-a + {\rm i}\RR_{<0}) \setminus B_r$.
 This property ensures that the notions of
 ``to the left of $C$'' and ``to the right of C''
 are well-defined.
 \item[$(2)$] For all $q$,
 the set $\alpha_q - \ZZ_{> 0} + \half$ is to the left of $C$,
 and the set $\alpha_q + \ZZ_{\geq 0} + \half$
 is to the right of $C$.
 \item[$(3)$] The path $C$ is symmetric about the origin,
 that is,
 for every complex number $z$ on $C$,
 $-z$ is on $C$.
\end{enumerate}

Because the parameter $c$ is assumed to be generic,
no two $\alpha_q$ have the same imaginary part,
so a curve satisfying condition~(2) exists.
Also, since $\alpha_q = -\alpha_{-q - 1}$ by Lemma~\ref{lem:a_prop},
the curve may be taken to also satisfy condition~(3).
Finally, since there are finitely many $\alpha_q$,
the curve may also be taken to satisfy condition~(1),
so a good contour is guaranteed to exist.

Define the polynomial
\begin{equation*}
 \bP(X) = \prod\limits_{q = 0}^{n - 1}
 \bigl( X - \exp\bigl(\trn \bigl( \alpha_q - q - \thalf \bigr)\bigr)
 \bigr),
\end{equation*}
so that $\bP\bigl(\exp\bigl(\trn z\bigr)\bigr) = 0$ when
$z \equiv \alpha_q - q - \half \pmod{n}$.
It follows from the definition that the only root of
$\bP\bigl(\exp\bigl(\rn \bigl(z - q - \half\bigr)\bigr)\bigr)$
between $-\half + C$ and $\half + C$
is at $z = \alpha_q$.
Moreover, the following lemma about $\bP(X)$ holds.

\begin{Lemma}
 \label{lem:P_sym}
 The polynomial $\bP(X)$ satisfies $\bP(X) = X^n \bP(X^{-1})$.
\end{Lemma}

\begin{proof}
 For each root $\exp\bigl(\rn \bigl(\alpha_q - q - \half\bigr)\bigr)$ of $\bP(X)$,
 the complex number
 \begin{align*}
 \exp\bigl(-\trn \bigl( \alpha_q - q - \thalf \bigr)\bigr)
 ={}& \exp\bigl(\trn \bigl( \alpha_{-q - 1} + q + \thalf \bigr)\bigr)= \exp\bigl(\trn
 \bigl( \alpha_{-q - 1} - (-q - 1) - \thalf \bigr)\bigr)
 \end{align*}
 is also a root of $\bP(X)$,
 so the reciprocal of each root of $\bP(X)$
 is also a root of $\bP(X)$.
 The claim follows.
\end{proof}

As the following theorem shows,
traces can be described as an integral over a good contour.

\begin{Theorem}
 \label{thm:analytic_formula}
 Define
 \begin{align*}
 w_0 (z) =
 \exp\bigl(\trn z\bigr) \cdot
 \frac{G\bigl(\exp\bigl(\rn z\bigr)\bigr)}{\bP\bigl(\exp\bigl(\rn z\bigr)\bigr)}, \qquad
 w_q (z) = w_0 (z - q),
 \end{align*}
 where $G$ is a polynomial of degree at most $n - 2$
 such that $G(X) = X^{n - 2} G\bigl(X^{-1}\bigr)$,
 and let $C$ be a good contour.
 All traces $T \colon H_c \rightarrow \CC$ are given by
 \begin{equation*}
 T ( R(k) e_q ) = \int_C R(z) w_q(z) \d z
 \end{equation*}
 on $\CC[k] e_q$, and extended to $H_c$
 using Corollary~{\rm\ref{cor:unique}}
 and the values of $T$ on the conjugacy classes~${\big\{g^{2i} h \mid i \in \ZZ\big\}}$ and $\big\{g^{2i + 1} h \mid i \in \ZZ\big\}$.
\end{Theorem}

\begin{proof}
 The weight functions $w_q$ satisfy the following properties:
 \begin{enumerate}[label=\numbers]\itemsep=0pt
 \item $w_q (z) = w_{q - 1} (z - 1)$,
 and so $w_q (z) = w_q (z - n)$;
 \item $w_q (z)$ decays exponentially
 when $\Im(z)$ approaches $\pm \infty$;
 \label{cond:exp}
 \item $(z - \alpha_q) w_q \bigl(z - \half\bigr)$ is holomorphic
 for $z$ between $-\half + C$ and $\half + C$.
 \label{cond:hol}
 \end{enumerate}
 Indeed, Property~\ref{cond:exp} holds
 because $\bP(X)$ is not divisible by $X$
 and $G$ has degree $\leq n - 2$,
 and Property~\ref{cond:hol} holds
 because the only root of $\bP\bigl(\exp\bigl(\rn \bigl(z - \half - q\bigr)\bigr)\bigr)$
 between $-\half + C$ and $\half + C$
 is at $z = \alpha_q$,
 and the function
 \[
 \frac{z - \alpha_q}{\exp\bigl(\rn z\bigr) - \exp\bigl(\rn \alpha_q\bigr)}\]
 is holomorphic at $z = \alpha_q$.

 We now check conditions~\ref{cond:shift}~and~\ref{cond:even}
 of Theorem~\ref{thm:classif}.
 First,
 \begin{gather*}
 T\bigl( \bigl( k + \thalf - \alpha_q \bigr)
 R\bigl( k + \thalf \bigr) e_q \bigr)
 - T\bigl( \bigl( k - \thalf - \alpha_q \bigr)
 R\bigl( k - \thalf \bigr) e_{q + 1} \bigr) \\
 \qquad= \int_{C} \bigl( z + \thalf - \alpha_q \bigr)
 R(z + \thalf) w_q(z) \d z
 - \int_{C} \bigl( z - \thalf - \alpha_q \bigr)
 R\bigl( z - \thalf \bigr) w_{q + 1}(z) \d z \\
 \qquad= \int_{\half + C} (z - \alpha_q) R(z)
 w_q \bigl( z - \thalf \bigr) \d z
 - \int_{-\half + C} (z - \alpha_q) R(z)
 w_{q + 1} \bigl( z + \thalf \bigr) \d z \\
 \qquad= \int_{\half + C} (z - \alpha_q) R(z)
 w_q \bigl( z - \thalf \bigr) \d z
 - \int_{-\half + C} (z - \alpha_q) R(z)
 w_q \bigl( z - \thalf \bigr) \d z.
 \end{gather*}
 Letting $U$ be the region between $-\half + C$ and $\half + C$,
 this is equal to
 \[
 \int_{\partial U}
 (z - \alpha_q) R(z) w_q \left( z - \thalf \right) \d z,
 \]
 which is equal to $0$ because
 $(z - \alpha_q) R(z) w_q \bigl(z - \half\bigr)$
 is holomorphic in $U$
 by Property~\ref{cond:hol} of $w_q$
 listed at the beginning of the proof.
 Thus condition~\ref{cond:shift} of Theorem~\ref{thm:classif} holds.

 To show that condition~\ref{cond:even} holds,
 we first show that $w_0$ is even.
 Indeed,
 \begin{align*}
 w_0 (z) ={}& \exp\bigl(\trn z\bigr) \cdot
 \frac{G\bigl( \exp\bigl(\rn z\bigr) \bigr)}
 {\bP\bigl( \exp\bigl(\rn z\bigr) \bigr)} = \exp\bigl(\trn z\bigr) \cdot
 \frac{\exp\bigl((n - 2) \cdot \rn z\bigr)
 G\bigl( \exp\bigl(-\rn z\bigr) \bigr)}
 {\exp\bigl(n \cdot \rn z\bigr)
 \bP\bigl( \exp\bigl(-\rn z\bigr) \bigr)} \\
 ={}& \exp\bigl(-\trn z\bigr) \cdot
 \frac{G\bigl( \exp\bigl(-\rn z\bigr) \bigr)}
 {\bP\bigl( \exp\bigl(-\rn z\bigr) \bigr)} = w_0 (-z).
 \end{align*}
 Recalling that $C$ is symmetric about the origin, we obtain
 \begin{align*}
 T( R(k) e_q )
 ={}& \int_{C} R(z) w_q(z) \d z = \int_{C} R(z) w_0 (z - q) \d z = \int_{C} R(-z) w_0 (-z - q) \d z \\
 ={}& \int_{C} R(-z) w_0 (z + q) \d z = \int_{C} R(-z) w_{-q}(z) \d z = T( R(-k) e_{-q} ),
 \end{align*}
 so condition~\ref{cond:even} of Theorem~\ref{thm:classif} holds.

 Finally, the dimension of the space of possible polynomials $G$
 is $\frac{n}{2}$ and distinct polynomials yield distinct traces,
 so, taking into account the value of the trace
 on the conjugacy classes~$\big\{g^{2i} h \mid i \in \ZZ\big\}$ and $\big\{g^{2i + 1} h \mid i \in \ZZ\big\}$,
 the dimension of the space of traces described by the theorem
 is $\frac{n}{2} + 2$, which is equal to the dimension
 of the space of all traces by Proposition~\ref{prop:dim}.
\end{proof}

The following theorem reduces the analytic formula
of Theorem~\ref{thm:analytic_formula}
to an integral over the complex line.

\begin{Theorem}
 \label{thm:analytic_formula_R}
 Define $w_q$ as in Theorem~{\rm\ref{thm:analytic_formula}}.
 All traces $T \colon H_c \rightarrow \CC$ are of the form
 \[
 T(R(k) e_q) = \int_{{\rm i}\RR} R(z) w_q(z) \d z
 + \Phi_q(R),
 \]
 where $\Phi_q$ is a linear functional of the form
 \[
 \Phi_q(R) = \sum_{\substack{i, j \in \ZZ, \\ i \neq q}}
 a_{q, i, j} R\bigl(\alpha_i + q - i + nj - \thalf\bigr)
 \]
 with finitely many $a_{q, i, j}$ not equal to zero.
 If the linear functional $\Phi_0$ is identically zero,
 then the trace $T$ is equivalent to a trace
 of the filtered deformation
 constructed from the polynomial~$P_{\circ}$
 obtained by removing roots of $P$ outside the strip
 $|{\Re(z)}| \leq \half$.
\end{Theorem}

\begin{proof}
 In the case when all $\alpha_q$ satisfy $|{\Re(\alpha_q)}| < \half$,
 the good contour $C$ may be taken to be~${\rm i}\RR$, proving the theorem.
 Otherwise, we have
 \begin{align*}
 T(R(k) e_q) ={}& \int_C R(z) w_q(z) \d z = \int_{{\rm i}\RR} R(z) w_q(z) \d z + \sum_{z_0} \Res_{z_0} (R w_q),
 \end{align*}
 where the sum is over all $z_0$ between ${\rm i}\RR$ and $C$
 where $w_q(z)$ has a singularity.
 In particular, these $z_0$ are of the form $\alpha_i + q - i + nj - \half$,
 with $i \in [0, n - 1]$ and $j \in \ZZ$.
 The value $\Res_{z_0} (R w_q)$ is proportional to $R(z_0)$,
 proving that the trace $T$ has the form claimed.

 If $\Phi_0$ is identically zero, then, for all polynomials $R$,
 the sum $\sum_{z_0} \Res_{z_0} (Rw_0)$
 over the poles~$z_0$ of $w_q(z)$ between ${\rm i}\RR$ and $C$
 is always zero.
 Taking $R$ to have arbitrary values on the poles
 (which is possible because the parameters $\alpha_q$ are generic,
 and hence the values $\alpha_i + q - i + nj$ are distinct),
 we find that all residues of these poles are zero.
 But this implies that there are no poles of $w_0$ between ${\rm i}\RR$ and $C$,
 which is only possible if, for all
 complex numbers $z_0$ of the form~${\alpha_i - i + nj - \half}$,
 the monomial $\bigl(z - \exp\bigl(\frac{2\pi {\rm i}}{n} z_0\bigr)\bigr)$ divides
 the polynomial $G$ of Theorem~\ref{thm:analytic_formula}.
 This monomial cancels out the corresponding divisor
 of the denominator $\bP$ of $w_0$,
 reducing the trace to one over $P_\circ$,
 as claimed.
\end{proof}

\begin{Remark}
 \label{rem:phi_0}
 In the case $q = 0$,
 the poles $\alpha_i - i + nj - \half$ are symmetric about the origin
 because $\alpha_i = -\alpha_{-i - 1}$ by Lemma~\ref{lem:a_prop}.
 As was shown in the proof of Theorem~\ref{thm:analytic_formula},
 $w_0$ is even, so~\smash{$\Res_{\alpha_i - i + nj - \half} (w_0) =
 \Res_{\alpha_{-i - 1} + i - nj + \half} (w_0)$}.
 It follows that $a_{0, i, j} = a_{0, -i - 1, -j}$.
\end{Remark}

\section{Positive traces}
\label{sec:pos}

In this section, we study positive traces on $\cO_c$
and prove our main result.
Recall that
the antilinear automorphism $\rho \colon H_c \rightarrow H_c$
is defined by
$\rho(x) = y$, $\rho(y) = -x$, $\rho(g) = g$, $\rho(h) = h$.
It follows that $\rho(k) = -k$ and $\rho(e_q) = e_{-q}$.

We use the following lemma.

\begin{Lemma}[{\cite[Lemma~4.2]{etingof2021twisted}}]
 \label{lem:analytic}
 Let $w(z)$ be a measurable nonnegative function on $\RR$
 such that~${w(z) < c \exp(-b|z|)}$
 for some positive constants $b, c$
 and which is positive almost everywhere.
 \begin{enumerate}[label=\numbers]\itemsep=0pt
 \item If $H(z)$ is a continuous complex-valued function on $\RR$
 with finitely many zeros
 and at most polynomial growth at infinity,
 then the set~$\{ H(z) S(z) \mid S(z) \in \CC[z] \}$
 is dense in the space~$L^p (\RR, w)$.
 \item The closure of the set
 $\{ S(z) \S(z) \mid S(z) \in \CC[z] \}$
 in $L^p (\RR, w)$ is the subset
 of almost everywhere nonnegative functions.
 \end{enumerate}
\end{Lemma}

We begin by showing that the linear functional $\Phi_0$
of Theorem~\ref{thm:analytic_formula_R}
must be identically zero.

\begin{Proposition}
 \label{prop:phi_0}
 Any positive trace $T$ on $\cO_c$
 of the form given by Theorem~{\rm\ref{thm:analytic_formula_R}}
 satisfies ${\Phi_0 (R) = 0}$ for all polynomials~$R$.
\end{Proposition}

\begin{proof}
 Since $T$ is a positive trace,
 we must have $T(a \rho(a)) > 0$ for all $a$ in $\cO_c$.
 In particular, we must have $T\big(R(k) \R(k) e\big) > 0$
 for all even polynomials $R$.
 By condition~\ref{cond:k_poly} of Theorem~\ref{thm:classif},
 this trace value is equal to
 $\half T\big(R(k) \R(k) e_0\big) + \half R(0) \R(0) T(e_0 h)$.

 Suppose, for the sake of contradiction,
 that $\Phi_0$ is not identically zero.
 Then there exists a~polynomial $S_0$
 such that $S_0(0) = 0$ and $\Phi_0(S_0)$ is negative.
 By Remark~\ref{rem:phi_0}, we may take $S_0$ to be even.
 Also, let
 \[
 H(z) = z \prod_{z_0} (z - z_0),
 \]
 where the product is over singularities $z_0$ of $w_0(z)$ between ${\rm i}\RR$ and the path $C$,
 so that $H(0) = 0$ and $H\bigl(\alpha_i - i - \half\bigr) = 0$
 for all $i$.
 Thus for any polynomial $U(z)$, we have
 $\Phi_0 \bigl( (S_0 - HU)\bigl(\S_0 - \bar{H} \bar{U}\bigr) \bigr) = \Phi_0(S_0)$
 and $S_0(0) - H(0) U(0) = 0$.
 Also, by Remark~\ref{rem:phi_0}, the polynomial $H(z)$ is even.

 By part (1) of Lemma~\ref{lem:analytic},
 there exists a sequence of polynomials $S_i$
 such that $HS_i$ tends to~$S_0$ in the space $L^2({\rm i}\RR, |w_0|)$.
 Because $w_0(z)$ is even,
 the sequence \smash{$H(z) \cdot \frac{S_i(z) + S_i(-z)}{2}$}
 also tends to $S_0$ in $L^2({\rm i}\RR, |w_0|)$.
 Thus the integral
 \[
 \int_{{\rm i}\RR} \left( S_0(z) - H(z) \cdot
 \frac{S_i(z) + S_i(-z)}{2} \right)
 \left( \S_0(z) - \bar{H}(z) \cdot
 \frac{\S_i(z) + \S_i(-z)}{2} \right)
 w_0(z) \d z
 \]
 tends to $0$.
 Letting $R(z) = S_0(z) - H(z) \cdot \frac{S_i(z) + S_i(-z)}{2}$
 for large enough $i$,
 we obtain
 \begin{align*}
 T\bigl(R(k) \R(k) e\bigr)
 ={}& \thalf T\bigl(R(k) \R(k) e_0\bigr) + \thalf R(0) \R(0) T(e_0 h) = \thalf T\bigl(R(k) \R(k) e_0\bigr) \\
 ={}& \int_{{\rm i}\RR} R(z) \R(z) w_0(z) \d z + \Phi_0 (R),
 \end{align*}
 which is negative, contradicting the positivity of the trace.
 Thus $\Phi_0$ must be identically zero, as claimed.
\end{proof}

We now prove a few preliminary results
which restrict a positive trace $T$
to be nonzero only on the span of terms of the form
$R(k) e_q$, with $R$ a polynomial.

\begin{Proposition}
 \label{prop:pos_e_0}
 Any positive trace $T$ on $\cO_c$ satisfies $T(e_0 h) = 0$.
\end{Proposition}

\begin{proof}
 Let $a$ = $R(k) (x^n + y^n) e$
 for some even polynomial $R$.
 The element $a = ea$ is in $\cO_c$, so by hypothesis,
 we must have $T(a \rho(a)) > 0$.
 Expanding and using the evenness of $R$ gives
 \begin{align*}
 T(a \rho(a)) ={}& T\bigl( R(k) (x^n + y^n) e
 \R(k) (x^n + y^n) e \bigr) \\
 ={}& T\bigl( R(k) (x^n + y^n)
 \R(k) (x^n + y^n) e \bigr) \\
 ={}& T\bigl( R(k) (x^n + y^n)
 \bigl( x^n \R(k - n) + y^n \R(k + n) \bigr)
 e \bigr),
 \end{align*}
 and distributing and using the commutativity relations
 with $k$ listed in Lemma~\ref{lem:k_prop},
 we obtain
 \begin{align*}
 T(a \rho(a))
 ={}& T\bigl( R(k) \bigl( x^{2n} \R(k - n) + x^n y^n \R(k + n)
 + y^{2n} \R(k + n) + y^n x^n \R(k - n) \bigr)
 e \bigr) \\
 ={}& T\bigl( x^{2n} R(k - 2n) \R(k - n) e \bigr)
 + T\bigl( x^n y^n R(k) \R(k + n) e \bigr) \\
 & + T\bigl( y^{2n} R(k + 2n) \R(k + n) e \bigr)
 + T\bigl( y^n x^n R(k) \R(k - n) e \bigr).
 \end{align*}
 Because $e_0$ is the average of the elements in $C_n$
 and $e$ is the average of the elements in $\DD_n$,
 we have $e = \frac{e_0 + e_0 h}{2}$.
 Expanding the expression for $T(a \rho(a))$
 and removing terms equal to $0$,
 we~obtain
 \begin{align*}
 T\bigl( a \rho(a) \bigr)
 ={}& \thalf T\bigl( x^{2n} R(k - 2n) \R(k - n) e_0 h \bigr)
 + \thalf T\bigl( x^n y^n R(k) \R(k + n) e_0 h \bigr) \\
 & + \thalf T\bigl( y^{2n} R(k + 2n) \R(k + n) e_0 h
 \bigr)
 + \thalf T\bigl( y^n x^n R(k) \R(k - n) e_0 h \bigr) \\
 & + \thalf T\bigl( x^n y^n R(k) \R(k + n) e_0
 \bigr)
 + \thalf T \bigl( y^n x^n R(k) \R(k - n) e_0 \bigr).
 \end{align*}
 Using conditions~\ref{cond:h_av} and~\ref{cond:h_av_flip}
 of Theorem~\ref{thm:classif}
 and rearranging terms, we obtain
 \begin{align*}
 T( a \rho(a) ) ={}& \thalf T\bigl( x^n y^n R(k - n) \R(k) e_0 h \bigr)
 + \thalf T\bigl( x^n y^n R(k) \R(k + n) e_0 h \bigr) \\
 & + \thalf T\bigl( y^n x^n R(k + n) \R(k) e_0 h \bigr)
 + \thalf T\bigl( y^n x^n R(k) \R(k - n) e_0 h \bigr) \\
 & + \thalf T\bigl( x^n y^n R(k) \R(k + n) e_0 \bigr)
 + \thalf T \bigl( y^n x^n R(k) \R(k - n) e_0 \bigr).
 \end{align*}
 Equating terms containing $y^n x^n$
 with terms containing $x^n y^n$
 using conjugation by $h$
 and combining like terms gives
 \begin{align}
 \label{eq:a_r_a}
 T\bigl( a \rho(a) \bigr)
 ={}& T\bigl( x^n y^n e_0 h \bigr) \bigl(R(n) \R(0) + R(0) \R(n)\bigr)
 + T\bigl( x^n y^n R(k) \R(k + n) e_0 \bigr) \nonumber \\
 ={}& T(e_0 h) \bigl(R(n) \R(0) + R(0) \R(n)\bigr)
 \prod\limits_{q = 0}^{n - 1} (c_0 q + \beta_q) \nonumber\\
 & + \int_{{\rm i}\RR} R(z) \R(z + n)
 \left( \prod\limits_{q = 0}^{n - 1}
 (c_0 (z + q) + \beta_q)\right) w_0 (z) |\d z|.
 \end{align}
 The product \smash{$\prod_{q = 0}^{n - 1} (c_0 q + \beta_q)$}
 in the first term is nonzero by assumption;
 indeed, by Lemma~\ref{lem:a_b_rel},
 $c_0 q + \beta_q = c_0 \bigl(q + \half - \alpha_q\bigr)$,
 which is equal to zero only if $\alpha_q - \half$ is an integer,
 which is impossible.

 Now assume, for the sake of contradiction,
 that $T(e_0 h)$ is nonzero.
 Letting $S(z)$ be the polynomial
 such that $R(z) = S\bigl(z - \frac{n}{2}\bigr)$
 and using the evenness of $R$,
 we obtain
 \begin{align*}
 T(a \rho(a)) ={}& T(e_0 h) \bigl(R(n) \R(0) + R(0) \R(n)\bigr)
 \prod\limits_{q = 0}^{n - 1} (c_0 q + \beta_q) \\
 & + \int_{{\rm i}\RR} R(-z) \R(z + n)
 \left( \prod\limits_{q = 0}^{n - 1}
 (c_0 (z + q) + \beta_q)\right) w_0 (z) |\d z| \\
 ={}& T(e_0 h) \bigl(S\bigl(\tfrac{n}{2}\bigr) \S\bigl(-\tfrac{n}{2}\bigr)
 + S\bigl(-\tfrac{n}{2}\bigr) \S\bigl(\tfrac{n}{2}\bigr)\bigr)
 \prod\limits_{q = 0}^{n - 1} (c_0 q + \beta_q) \\
 & + \int_{{\rm i}\RR}
 S\bigl(-z - \tfrac{n}{2}\bigr) \S\bigl(z + \tfrac{n}{2}\bigr)
 \left( \prod\limits_{q = 0}^{n - 1}
 (c_0 (z + q) + \beta_q)\right) w_0 (z) |\d z| \\
 ={}& T(e_0 h) \bigl(S\bigl(\tfrac{n}{2}\bigr) \S\bigl(-\tfrac{n}{2}\bigr)
 + S\bigl(-\tfrac{n}{2}\bigr) \S\bigl(\tfrac{n}{2}\bigr)\bigr)
 \prod\limits_{q = 0}^{n - 1} (c_0 q + \beta_q) \\
 & + \int_{{\rm i}\RR + \frac{n}{2}} S(-z) \S(z) w(z) |\d z|,
 \end{align*}
 where $w$ is a function with
 $|w(z)|$ satisfying the conditions of Lemma~\ref{lem:analytic}\,(1).
 Because $\beta_q = \frac{c_0}{2} - c_0 \alpha_q$
 by Lemma~\ref{lem:a_b_rel},
 the weight function $w(z)$ is holomorphic
 between ${\rm i}\RR$ and ${\rm i}\RR + n$.
 Moving the contour of integration, we obtain
 \begin{equation*}
 T(a \rho(a)) = \psi(S)
 + \int_{{\rm i}\RR} S(-z) \S(z) w(z) |\d z|,
 \end{equation*}
 where $\psi(S)$ is proportional to
 $S(\frac{n}{2})\S\bigl(-\frac{n}{2}\bigr) + S\bigl(-\frac{n}{2}\bigr)\S\bigl(\frac{n}{2}\bigr)$.
 By suitable choice of $S$,
 we can force~$\psi(S)$ to be negative;
 let $S_0$ be a polynomial with this property
 and satisfying that~$S_0(z) = S_0(n - z)$,
 and let \smash{$U(z) = \bigl(z - \frac{n}{2}\bigr)^2$},
 so that $\psi(S_0 - UL) = \psi(S_0)$ for any polynomial $L$
 and~$U(z) = U(n - z)$.
 By Lemma~\ref{lem:analytic}\,(1), there exists a sequence of polynomials
 $S_i$ such that $U S_i$ tends to $S_0$
 in the space $L^2 ({\rm i}\RR, |w(z)| + |w(n - z)|)$.
 It follows from the symmetry of $S_0$ and $U$ that~\smash{$U(z) \cdot \frac{S_i(z) + S_i(n - z)}{2}$}
 tends to $S_0$
 in the space $L^2({\rm i}\RR, |w(z)|)$.
 Therefore, the integral
 \begin{gather*}
 \int_{{\rm i}\RR} \left( S_0 (-z)
 - U(-z) \cdot \frac{S_i (-z) + S_i (n + z)}{2} \right)\\
 \qquad\times
 \left( \S_0 (z)
 - \bar{U}(z) \cdot \frac{\S_i (z) + \S_i (n - z)}{2}
 \right)
 w(z) |\d z|
 \end{gather*}
 approaches $0$.
 On the other hand,
 \smash{$\psi\bigl(S_0 (z) - U(z) \cdot \frac{S_i (z) + S_i (n - z)}{2}\bigr)
 = \psi(S_0)$}
 is negative.
 Therefore, choosing
 \smash{$S(z) = S_0 (z) - U(z) \cdot \frac{S_i (z) + S_i (n - z)}{2}$}
 for large enough $i$
 and noting that the resulting polynomial $R$ is even,
 we obtain $T(a \rho(a)) < 0$,
 a contradiction.
\end{proof}

We now obtain a similar result for $e_m h$.

\begin{Proposition}
 \label{prop:pos_e_m}
 Any positive trace $T$ on $\cO_c$ satisfies $T(e_m h) = 0$.
\end{Proposition}

\begin{proof}
 Let $a = (x^n + y^n + S(k))e$ for some even polynomial $S$.
 As in Proposition~\ref{prop:pos_e_0}, $ea = a$, so $a$ is in $\cO_c$,
 and we must have $T(a \rho(a)) = 0$.
 Expanding and using the evenness of $S$ gives
 \begin{align*}
 T(a \rho(a)) ={}& T\bigl( (x^n + y^n + S(k)) e
 \bigl(x^n + y^n + \S(k)\bigr) e \bigr) \\
 ={}& T\bigl( (x^n + y^n + S(k)) (x^n + y^n + \S(k)) e \bigr) \\
 ={}& T\bigl( \bigl(x^{2n} + x^n y^n + y^n x^n + y^{2n}\bigr) e \bigr)
 + T\bigl( S(k) \S(k) e \bigr) \\
 & + T\bigl( x^n \bigl(S(k - n) + \S(k)\bigr) e \bigr)
 + T\bigl( y^n (S(k + n) + \S(k)) e \bigr).
 \end{align*}
 We evaluate these terms separately.
 By Proposition~\ref{prop:pos_e_0} and \eqref{eq:a_r_a} with $R = 1$, we have
 \begin{align*}
 T\bigl( (x^{2n} + x^n y^n + y^n x^n + y^{2n}) e \bigr)
 ={}& \int_{{\rm i}\RR} \prod\limits_{q = 0}^{n - 1}
 (c_0 (z + q) + \beta_q) w_0 (z) |\d z|.
 \end{align*}
 Splitting into $e_0$ and $e_0 h$ components
 and using Proposition~\ref{prop:pos_e_0},
 we find that the second term is simply
 \begin{align*}
 T\bigl( S(k) \S(k) e \bigr)
 ={}& \thalf T\bigl( S(k) \S(k) e_0 \bigr)
 + \thalf T\bigl( S(k) \S(k) e_0 h \bigr)
 = \int_{{\rm i}\RR} S(z) \S(z) w_0 (z) |\d z|.
 \end{align*}
 Again using Proposition~\ref{prop:pos_e_0}
 and applying conditions~\ref{cond:h_av} and~\ref{cond:h_av_flip}
 of Theorem~\ref{thm:classif},
 we find that
 the sum of the last two terms is equal to
 \begin{gather*}
 \thalf T\bigl( x^n (S(k - n) + \S(k)) e_0 h \bigr)
 + \thalf T\bigl( y^n (S(k + n) + \S(k)) e_0 h \bigr) \\
 \qquad= \thalf T\bigl( x^m y^m (S(k - m) + \S(k + m)) e_m h \bigr)\\
 \phantom{ \qquad= }{}
 + (-1)^m \thalf T\bigl( y^m x^m (S(k + m) + \S(k - m))
 e_m h \bigr),
 \end{gather*}
 which, by conjugation by $h$, is equal to
 \begin{equation*}
 T\bigl(x^m y^m \bigl(S(k - m) + \S(k + m)\bigr) e_m h\bigr)
 = T(e_m h) \bigl(S(-m ) + \S(m)\bigr)
 \prod\limits_{q = 0}^{m - 1} (c_0 q + \beta_{m + q}).
 \end{equation*}
 Thus
 \begin{align*}
 T(a \rho(a)) ={}& T(e_m h) \bigl(S(-m ) + \S(m)\bigr)
 \prod\limits_{q = 0}^{m - 1} (c_0 q + \beta_{m + q}) + \int_{{\rm i}\RR} \prod\limits_{q = 0}^{n - 1}
 (c_0 (z + q) + \beta_q) w_0 (z) |\d z| \\
 & + \int_{{\rm i}\RR} S(-z) \S(z) w_0 (z) |\d z|.
 \end{align*}
 Assume, for the sake of contradiction, that $T(e_m h)$
 is nonzero.
 As in the proof of Proposition~\ref{prop:pos_e_0},
 the product $\prod_{q = 0}^{m - 1} (c_0 q + \beta_{m + q})$
 is nonzero by Lemma~\ref{lem:a_b_rel}
 and the condition on the $\alpha_q$,
 so for a~suitable choice $S_0$ of even $S$,
 the expression
 \begin{equation*}
 T(e_m h) \bigl(S(-m ) + \S(m)\bigr)
 \prod\limits_{q = 0}^{m - 1} (c_0 q + \beta_{m + q})
 + \int_{{\rm i}\RR} \prod\limits_{q = 0}^{n - 1}
 (c_0 (z + q) + \beta_q) w_0 (z) |\d z|
 \end{equation*}
 can be made negative.
 Taking $U(z) = (z - m)(z + m)$,
 there exists a sequence $S_i$ of polynomials
 such that $US_i$ approaches $S_0$ in
 $L^2 ({\rm i}\RR, |w_0 (z)|)$
 by Lemma~\ref{lem:analytic}\,(1),
 so \smash{$U(z) \cdot \frac{S_i (z) + S_i (-z)}{2}$}
 approaches $S_0$ in $L^2 ({\rm i}\RR, |w_0 (z)|)$
 because $w_0$ is even.
 Setting
 \[
 S(z) = S_0(z) - U(z) \cdot \frac{S_i(z) + S_i(-z)}{2}
 \]
 for large enough $i$,
 it follows that $T(a \rho(a)) < 0$, a contradiction.
\end{proof}

We now obtain our main result.

\begin{Theorem}
 \label{thm:pos_trace}
 Let $c$ be generic and contained in $\CC[C_n]$.
 Any positive trace of a filtered deformation $\cO_c^{\DD_n}$
 of a Kleinian singularity of type D
 is the restriction of a positive trace
 on the corresponding filtered deformation $\cO_c^{C_n}$
 of a Kleinian singularity of type A.
\end{Theorem}

\begin{proof}
 We first describe the correspondence between
 $\cO_c^{\DD_n}$ and $\cO_c^{C_n}$.
 Just as the algebra $\cO_c^{\DD_n}$
 is defined as $e (\CC[x, y] \# \DD_n) e$,
 a subalgebra of $H_c$,
 the algebra $\cO_c^{C_n}$
 is defined as $e_0 (\CC[x, y] \# C_n) e_0$,
 also a subalgebra of $H_c$.
 The elements of $\cO_c^{\DD_n}$
 are of the form $R(x, y) e$,
 with $R(x, y)$ a polynomial expression in $x$ and $y$
 invariant under the action of $\DD_n$,
 and the elements of $\cO_c^{C_n}$
 are of the form~$S(x, y) e_0$,
 with $S(x, y)$ a polynomial expression
 invariant under the action of $C_n$.
 The map~${R(x, y) e \mapsto R(x, y) e_0}$
 for invariant polynomials $R(x, y)$
 is an injective algebra homomorphism
 from $\cO_c^{\DD_n}$ to $\cO_c^{C_n}$,
 so we may identify $\cO_c^{\DD_n}$
 with a subalgebra of $\cO_c^{C_n}$.
 In particular,
 $\cO_c^{\DD_n}$~is mapped to the set of elements of $\cO_c^{C_n}$
 invariant under the involution given by
 $k e_0 \mapsto -k e_0$,
 $x^n e_0 \mapsto y^n e_0$,
 $y^n e_q \mapsto x^n e_0$.

 By this inclusion, any trace on $\cO_c^{C_n}$
 can be restricted to a trace on $\cO_c^{\DD_n}$.
 Observing that the formula in Theorem~\ref{thm:analytic_formula}
 is identical to the analytic formula for type A traces
 obtained in~\cite[Proposition~3.1]{etingof2021twisted}
 with \smash{$z = \frac{k}{n} e_0$}
 and $u$ and $v$ scalar multiples of
 \smash{$\frac{x^n}{c_0 n} e_0$} and \smash{$\frac{y^n}{c_0 n} e_0$},
 respectively,
 such that the polynomial $P$ satisfying $uv = P\bigl(z + \half\bigr)$
 obtained from Lemma~\ref{lem:xy_poly}
 is monic,
 we see that the resulting restricted trace $T$
 can be any trace satisfying Theorem~\ref{thm:analytic_formula}
 and the additional condition that $T(e_q h) = 0$
 for all $q$.
 This condition follows from
 Lemma~\ref{lem:trace_eh_zero} and Propositions~\ref{prop:pos_e_0} and~\ref{prop:pos_e_m},
 so any positive trace on $\cO_c^{\DD_n}$
 is the restriction of some trace on $\cO_c^{C_n}$.

 Let $T$ be a positive trace on $\cO_c^{\DD_n}$,
 and consider its unique even extension to $\cO_c^{C_n}$
 given by the even weight function $w_0$
 and the analytic formula
 in~\cite[Proposition~3.1]{etingof2021twisted}.
 We wish to show that~$T$ is positive on $\cO_c^{C_n}$.
 Because $T$ is positive on $\cO_c^{\DD_n}$,
 we must have $T(e_0 R(k) e_0 \rho(e_0 R(k) e_0)) > 0$
 for any even polynomial $R$,
 or equivalently
 \begin{equation}
 \label{eq:pos_w}
 \int_{{\rm i}\RR} R(z) \R(-z) w_0(z) |\d z| > 0.
 \end{equation}
 Also, letting $P$ be the polynomial such that
 $x^n y^n e_0 = P\bigl(k + \frac{n}{2}\bigr) e_0$,
 we have
 \begin{equation}
 \label{eq:pos_pw}
 \int_{{\rm i}\RR} R(z) \R(-z - n) P\bigl(z + \tfrac{n}{2}\bigr) w_0(z) |\d z|
 > 0
 \end{equation}
 for all even polynomials $R$ by \eqref{eq:a_r_a}.
 By Lemma~\ref{lem:analytic}\,(2),
 \eqref{eq:pos_w} is equivalent to $w_0 (z)$
 being nonnegative on ${\rm i}\RR$ because
 for a sequence of polynomials $S_i$ tending to
 some almost everywhere nonnegative even function $U$
 in $L^1(\RR, |w_0(z)|)$,
 the sequence of polynomials $\frac{S_i(z) + S_i(-z)}{2}$
 tends to $U$ in $L^1(\RR, |w_0(z)|)$
 because $w_0$ is even.

 Also, letting $S$ be the polynomial such that
 $R(z) = S(z + \frac{n}{2})$
 and noting that $P(z + \frac{n}{2}) w_0(z)$ is holomorphic
 between ${\rm i}\RR$ and $\frac{n}{2} + {\rm i}\RR$,
 \eqref{eq:pos_pw} is equivalent to the statement that
 \begin{equation*}
 \int_{\frac{n}{2} + {\rm i}\RR}
 S(z) \S(-z) P(z) w_0 \bigl(z - \tfrac{n}{2}\bigr) |\d z| > 0
 \end{equation*}
 for all polynomials $S$ such that $S(z) = S(n - z)$.
 Again using Lemma~\ref{lem:analytic}\,(2),
 it follows that~${P(z) w_0\bigl(z - \frac{n}{2}\bigr)}$ is nonnegative
 on ${\rm i}\RR$.

 The positivity conditions obtained on $P$ and $w_0$
 are equivalent to those obtained
 in~\cite[Proposition~4.20]{etingof2021twisted} for type A,
 so the trace $T$ is positive on~$\cO_c^{\DD_n}$
 if and only if it is positive on~$\cO_c^{C_n}$.
 Thus $T$ is the restriction of a positive trace on $\cO_c^{C_n}$,
 as desired.
\end{proof}

\pdfbookmark[1]{Part 2: Part 2: Pure SL(2)/PGL(2) gauge theory}{part2}
\begin{center}
 {\huge \bf Part 2: Pure $\boldsymbol{\SL(2)/\PGL(2)}$ gauge theory}
\end{center}

\section{Description of the algebra}
\label{SecAlgebra}

\subsection[Definition of A, connection to the BFN construction]{Definition of $\boldsymbol{\mc{A}}$, connection to the BFN construction}

 Let $0<q<1$ be a number. Consider the algebra $\mc{B}$ with generators $v$, $v^{-1}$, $u_{\pm}$ and relations~${u_{\pm }v=q^{\pm 1}vu_{\pm}}$, \[u_+u_-=\frac{1}{\bigl(v-v^{-1}\bigr)\bigl(qv-(qv)^{-1}\bigr)},\qquad u_-u_+=\frac{1}{\bigl(v-v^{-1}\bigr)\bigl(q^{-1}v-qv^{-1}\bigr)}.\]

\begin{Definition}
\label{DefOfA}
 $\mc{A}$ is the subalgebra of $\mc{B}$ generated by elements $v^k+v^{-k}$ and $H_a=q^{\tfrac a2}(v^au_++v^{-a}u_-)$ for all positive integers $k$ and all integers $a$.
\end{Definition}

Gaiotto and Teschner~\cite{GT} use $\mc{A}$ to describe the $K$-theoretic Coulomb branch of pure $\SL(2)$ and $\PGL(2)$ gauge theories. We will explain the connection from a mathematical point of view below.

First, we note that $\mc{A}$ can be embedded into a localized $q$-Weyl algebra.

\begin{Definition}

 Let $W_0=\CC\langle v,w\rangle/(wv-qvw)$ be a $q$-Weyl algebra. Then elements $v$, $w$ and $\big\{q^iv-q^{-i}{v^{-1}}\big\}$ generate a multiplicative subset $S$ satisfying both the left and the right Ore condition. We define $W$ to be the Ore localization of $W$ at $S$.
\end{Definition}
\begin{Lemma}
\label{LemBIntoW}
 There is an embedding of $\mc{B}$ into $W$ given by $v\mapsto v$, \smash{$u_+\mapsto w\bigl(v-v^{-1}\bigr)^{-1}$} and \smash{$u_-\mapsto w^{-1}\bigl(v-v^{-1}\bigr)^{-1}$}.
\end{Lemma}
\begin{proof}
 We see that the images satisfy the defining relations $u_{\pm}v=q^{\pm 1}uv_{\pm}$ of $\mc{B}$. For the other two relations, we have \[u_+u_-=w\bigl(v-v^{-1}\bigr)^{-1}w^{-1}\bigl(v-v^{-1}\bigr)^{-1}=\bigl(qv-q^{-1}v^{-1}\bigr)^{-1}\bigl(v-v^{-1}\bigr)^{-1},\]
and
\[u_-u_+=w^{-1}\bigl(v-v^{-1}\bigr)^{-1}w\bigl(v-v^{-1}\bigr)^{-1}=\bigl(q^{-1}v-qv^{-1}\bigr)^{-1}\bigl(v-v^{-1}\bigr)^{-1}.\] The basis of $\mc{B}$ is given by expressions~$v^l$, $u_+^kv^l$ and $u_-^k v^l$ for positive integers $k$ and integers $v_l$. Each $k$ corresponds to its own power of $w$, then different $l$ give linearly independent elements.
\end{proof}

We identify $\mc{A}$ with its image in $W$. This is the only property of $\mc{A}$ that we will use in Sections~\ref{SecTwistedTraces} and~\ref{SecPositiveTraces}. We discuss the connection to the BFN construction for completeness.

We turn to BFN Coulomb branches. Pure gauge theory means that $N=0$. In this case, BFN construction gives equivariant $K$-theory of the affine Grassmannian \[\mc{A}^q_{G,N}=K^{G(\CC[[t]])\rtimes \CC_q^{\times}}(\Gr_G).\] The group $\CC_q^{\times}$ acts by loop rotation on $G(O)$ and on $\Gr_G$.

Let $q$ be a formal parameter instead of a number, the definition of $\mc{A}$ works with the following change: $\mc{A}$ is a $\CC\big[q,q^{-1}\big]$-algebra and we change the generator $H_a$ to $\tilde{H}_a=(v^au_++v^{-a}u_-)$. Then~$\mc{A}$ should be equal to \smash{$K^{\SL(2,\CC[[t]])\rtimes \CC_q^{\times}}(\Gr_{\PGL_2})$}, a modification of BFN construction. We use this as a motivation to consider $\mc{A}$, but a rigorous proof is beyond the scope of this paper.

\begin{lemma*}
 We have \smash{$\mc{A}\cong K^{\SL(2,\CC[[t]])\rtimes \CC_q^{\times}}(\Gr_{\PGL_2})$}. More precisely, we identify $W$ with $K^{\CC^{\times}_v\times \CC^{\times}_q}\allowbreak\times(\Gr_T)_{\rm loc}$, the localized quantized Coulomb branch of pure $\CC^{\times}_v$ gauge theory. Then the image of the localization map
 \[(i_*)^{-1}\colon\ K^{\SL(2,\CC[[t]])\rtimes \CC_q^{\times}}(\Gr_{\PGL_2})\to K^{\CC^{\times}_v\times \CC^{\times}_q}(\Gr_T)_{\rm loc}\]
 equals to the image of $\mc{A}$ in $W$ under the map \smash{$u_+\mapsto \bigl(v-v^{-1}\bigr)^{-1}w$}, \smash{$u_-\mapsto \bigl(v-v^{-1}\bigr)^{-1}w^{-1}$}.
\end{lemma*}
\begin{proof}[Sketch of the proof]
 Note that we switched between left and right fractions compared to Lemma~\ref{LemBIntoW}, but the proof of the inclusion $\mc{B}\hookrightarrow W$ works without change.

 Similarly to the proof of \cite[Lemma~6.9]{BFN}, one can prove that \smash{$K^{\SL(2,\CC[[t]])\rtimes \CC_q^{\times}}(\Gr_{\PGL_2})$} is generated by the dressed miniscule monopole operators and the K-theory of a point. Then we use the localization map, it is also called the abelianization map in physics literature. For pure gauge theory this map is constructed in \cite[Section~6.3]{BFM}. We get the map
 \[
 (i_*)^{-1}\colon \ K^{\SL(2,\CC[[t]])\rtimes \CC_q^{\times}}(\Gr_{\PGL_2})\to K^{\CC^{\times}_v\times \CC^{\times}_q}(\Gr_T)_{\rm loc},
 \]
 where $\CC^{\times}_v\subset \SL(2,\CC)$ is the maximal torus and the multiplicative subset on the right is defined so that \smash{$W=K^{\CC^{\times}_v\times \CC^{\times}_q}(\Gr_T)_{\rm loc}$}. Let $v$ be the generator of the $\CC^{\times}_v$-equivariant $K$-theory of a~point corresponding to the standard representation, similarly for $q$ and $\CC^{\times}_q$. Then the image of \smash{$K^{\SL(2)}(pt)$} under the localization map consists of Laurent polynomials in $v$ symmetric with respect to $v\mapsto v^{-1}$. Dressed miniscule monopole operators are defined as $K$-theory classes of the sheaves $O(l)$, $l\in \ZZ$ on the smooth $\PGL_2(\CC[[t]])$-orbit $\mathbb{P}^1\subset \Gr_{\PGL_2}$. Here the $\SL_2$-equivariant structure is the natural one, and the action of $\CC_q^{\times}$ on $O(l)$ is trivial. It can be checked that the image of $[O(l)]$ in $W$ is
 \[
 -\bigl(v-v^{-1}\bigr)^{-1}\bigl(v^{l+1}w-v^{-l-1}w^{-1}\bigr)=-\tilde{H}_{l+1}.
 \]
 This finishes the proof.
\end{proof}

We claim that \smash{$K^{\SL(2,\CC[[t]])\rtimes \CC_q^{\times}}(\Gr_{\PGL_2})$} contains Coulomb branches of pure $\SL(2)$ and $\PGL(2)$ gauge theories. The inclusion \[\mc{A}_{\SL_2,0}^q=K^{\SL(2,\CC[[t]])\rtimes \CC_q^{\times}}(\Gr_{\SL_2})\subset K^{\SL(2,\CC[[t]])\rtimes \CC_q^{\times}}(\Gr_{\PGL_2})\] comes from the inclusion of the affine Grassmannians. Let $\tau_1\colon \mc{A}\to\mc{A}$ be the map that fixes $v$ and sends $w$ to $-w$. Then $\mc{A}_{\SL_2,0}^q$ is the subalgebra $\mc{A}^{\tau_1}$ of $\tau_1$-fixed elements. The inclusion \[\mc{A}_{\PGL_2,0}^q=K^{\PGL(2,\CC[[t]])\rtimes \CC_q^{\times}}(\Gr_{\PGL_2})\subset K^{\SL(2,\CC[[t]])\rtimes \CC_q^{\times}}(\Gr_{\PGL_2})\] follows from the fact that any $\PGL_2$-equivariant sheaf is $\SL_2$-equivariant via the projection $\SL_2\twoheadrightarrow \PGL_2$. In terms of the generators above, $\mc{A}_{\PGL_2,0}^q$ is generated by $v^n+v^{-n}$ with $n$ even and $H_a$ with $a$ odd. Any such element is fixed by the involution $\tau_2$ such that $\tau_2(v)=-v$, $\tau_2(w)=-w$ and we have $\mc{A}^{q}_{\PGL_2,0}=\mc{A}^{\tau_2}$.

Gaiotto and Teschner consider the automorphism $\rho(H_a)=H_{a-2}$ and $g(H_a)=H_{a-4}$, but the definition of twisted and positive trace in physics is slightly different: $T(ab)=T(g(b)a)$ instead of $T(ab)=T(bg(a))$ and $T(\rho(a)a)>0$ instead of $T(a\rho(a))>0$. Any twisted trace satisfies~${T(a)=T(g(a))}$. Hence, a $g$-twisted trace in one sense is a $g^{-1}$-twisted trace in another. So we consider $g(H_a)=H_{a+4}$ and, more generally, $g(H_a)=H_{a+m}$ and \smash{$\rho(H_a)=H_{a+\tfrac m2}$} for even~$m$.

We classify twisted and positive traces on $\mc{A}$ in the case when $m\geq 0$.

For future use, we note that $g$ and $\rho$ can be extended to an (anti-)automorphism of $W$ as follows:
 \begin{alignat*}{3}
 & g(v)=v,\qquad&& g(w)=q^{\frac{m}{2}}v^mw=q^{-\frac{m}{2}}wv^{m},&\\
 & \rho(v)=v^{-1},\qquad && \rho(w)=q^{\frac{m}{4}}v^{-\frac{m}{2}}w^{-1}=q^{-\frac{m}{4}} w^{-1}v^{-\frac{m}{2}}.&
\end{alignat*}

\begin{Remark}
 There is an isomorphism between $\mc{A}_q$ and $\mc{A}_{q^{-1}}$ that does not change $w$ and sends~$v$ to $v^{-1}$. This isomorphism interchanges $g_m$-twisted traces for $\mc{A}_q$ and $g_{-m}$-twisted traces for~$\mc{A}_{q^{-1}}$. It can be shown that in the integral formula from Proposition~\ref{PropSIsAnINtegral} we can take $q$ to be a~formal parameter and take formal power series $\omega(z)$ instead of an actual quasi-periodic function. This would give a classification of twisted traces for all $q$ and hence for all $m$. For~${0<q<1}$ and negative $m$, most of these power series will have quadratic exponential growth and converge nowhere, similarly to the series for Jacobi theta function $\sum q^{n^2}z^n$ with $q>1$ instead of $0<q<1$. Because of this and the fact that $m<0$ case is less physically significant, we work only with the case $m\geq 0$.
\end{Remark}
\begin{Remark}
 Geometrically, $g_m$ should correspond to the twist by $m$-th power of the determinant bundle on the affine Grassmannian.
\end{Remark}

\subsection[Description of the image A subset W]{Description of the image $\boldsymbol{\mc{A}\subset W}$}

We will write elements of $\mc{A}$ and $W$ as \smash{$a=\sum w^i f_i=\sum_{i=-N}^N w^i f_i$}, where $N$ is a nonnegative integer and $f_i$ is a rational function in $v$. The function $f_i$ may have poles only at $\pm q^k$ for some~$k$. We denote the residue of $f_i$ at $\pm q^k$ by \smash{$f_i^{(k)}$} or \smash{$f_i^{(k,-)}$}, respectively. We will also use notation the~\smash{$f_i^{(k,+)}=f_i^{(k)}$}
\begin{Proposition}
\label{PropAInsideW}
The algebra $\mc{A}$ coincides with the set of elements $a=\sum w^i f_i$ of $W$ such that all $f_i$ have simple poles only possibly at $\pm q^j$, $j\in\ZN$ and the corresponding residues \smash{$f_i^{(j,\pm)}$} satisfy
\begin{enumerate}\itemsep=0pt
 \item[$(1)$]
 \smash{$f_{-i}\bigl(v^{-1}\bigr)=(-1)^i f_i(v)$};
 \item[$(2)$]
 $f_i^{(i,\pm)}=0$;
 \item[$(3)$]
 for a positive integer $k$, we have \smash{$f_{i+k}^{(i,\pm)}=(-1)^{k+1}f_{i-k}^{(i,\pm)}$}, where $\pm$ is the same sign on both sides.
\end{enumerate}
In particular, \smash{$f_0^{(0,\pm)}=0$}, so that $f_0(v)$ has no poles at the unit circle.
\end{Proposition}
\begin{proof}
We should check that any element of $\mc{A}$ satisfies the conditions (1)--(3) and that any element satisfying conditions (1)--(3) belongs to $\mc{A}$.

 {\bf Elements of $\mc{A}$ satisfy the conditions.}
 Note that an element $a=\sum w^i f_i\in W$ satisfies the condition $f_{-i}\bigl(v^{-1}\bigr)=(-1)^i f_i(v)$ if and only if $a$ is fixed by the involution $\tau$ that sends $v$ to~$v^{-1}$ and $w$ to $-w^{-1}$. In terms of $u_{\pm}$, we have $\tau(u_+)=u_-$ and $\tau(u_-)=u_+$, so that $\tau(H_a)=H_a$. It follows that all generators of $\mc{A}$ are fixed by $\tau$. This shows that the first condition is satisfied.

 We check the second and third conditions by induction on the number of $H_a$ used in the multiplication. The base case is zero and one multiplications: $p(v)$ and \[
 H_a=q^{\frac{a}{2}}(v^au_+ + v^{-a}u_-)=\smash{q^{\frac{-a}{2}}(u_+v^a+u_-v^{-a})}.\]
 The polynomial $p(v)$ has no poles. For $H_a$, we write
 \[
 u_+v^a+u_-v^{-a}=\smash{wv^a\bigl(v-v^{-1}\bigr)^{-1}+w^{-1}v^{-a}\bigl(v-v^{-1}\bigr)^{-1}},
 \]
 so that \smash{$f_1=\frac{v^a}{v-v^{-1}}$} and \smash{$f_{-1}=\frac{v^{-a}}{v-v^{-1}}$}. All \smash{$f_i^{(j)}$} are zero except
 \[
 f_1^{(0)}=\tfrac{1}{2} , \qquad f_1^{(0,-)}=\tfrac{(-1)^a}{2} , \qquad f_{-1}^{(0)}=\tfrac{1}{2} , \qquad f_1^{(0,-)}=\tfrac{(-1)^{-a}}{2}.
 \]
 It follows that both the second and third condition are satisfied.

Note that the second and the third condition can be written as \smash{$f_{i+k}^{(i,\pm)}=(-1)^{k+1}f_{i-k}^{(i,\pm)}$}, where~$k$ is a nonnegative integer. For $k=0$ we get the second condition.

We turn to the induction step. We will prove that if $x=\sum w^i f_i$ satisfies the conditions then
\[
H_ax=(wg_1+w^{-1}g_{-1})x
\] also satisfies the second and third conditions. Here $g_{\pm 1}=\frac{v^{\pm a}}{v-v^{-1}}$ are rational functions in~$v$ with~at most simple pole at $v=\pm 1$ satisfying \smash{$g_{1}^{(1,\pm)}=g_{-1}^{(1,\pm)}$}. We~also have \smash{$g_{-1}=-g_1\bigl(v^{-1}\bigr)$}.

We have
\begin{align}
 H_ax&=\bigl(wg_1+w^{-1}g_{-1}\bigr)\Big(\sum w^i f_i\Big)=\sum w^i \bigl(g_1\bigl(q^{1-i}v\bigr)f_{i-1}+g_{-1}\bigl(q^{-1-i}v\bigr)f_{i+1}\bigr)\nonumber\\
 &=\sum w^i h_i. \label{EqHax}
\end{align}

Note that $g_{\pm 1}\bigl(q^{-i}v\bigr)$ has at most simple poles at $\pm q^{i}$ with residue \smash{$q^i g_{\pm 1}^{(0,\pm)}$}. Using the fact that $f_i$ is regular at $q^i$, we see that $g_{\pm 1}\bigl(q^{-i}v\bigr)f_i$ has simple poles.

Hence, $h_i$ has simple poles. It remains to prove that for all $i$ and $k\geq 0$, we have $\smash{h_{i+k}^{(i,\pm)}}=\smash{(-1)^{k+1}h_{i-k}^{(i,\pm)}}$. We will prove it for $\pm=+$, and for $\pm=-$ the proof is the same.

For $j\neq i\pm 1$, the residue \smash{$h_i^{(j)}$} can be computed as \begin{equation}
 \label{EqHij}
 h_i^{(j)}=g_1\bigl(q^{j+1-i}\bigr)f_{i-1}^{(j)}+g_{-1}\bigl(q^{j-1-i}\bigr)f_{i+1}^{(j)}.
\end{equation}

Hence, for $k\neq 1$, we have
\begin{align*}
 h_{i+k}^{(i)}&=g_1\bigl(q^{k+1}\bigr)f_{i+k-1}^{(i)}+g_{-1}\bigl(q^{k-1}\bigr)f_{i+k+1}^{(i)}\\
 &=-g_{-1}\bigl(q^{-k-1}\bigr)(-1)^k f_{i-k+1}^{(i)}-g_1\bigl(q^{1-k}\bigr)(-1)^kf_{i-k-1}^{(i)}=(-1)^{k+1}h_{i-k}^{(i)}.
\end{align*}

For $j=i+1$, we have \[h_i^{(i+1)}=g_1\bigl(q^2\bigr)f_{i-1}^{(i+1)}+q^{i+1}g_{-1}^{(0)}f_{i+1}\bigl(q^{i+1}\bigr),\]
and for $j=i-1$ we have
\begin{equation}
\label{EqHii1}
 h_i^{(i-1)}=q^{i-1}g_1^{(0)}f_{i-1}\bigl(q^{i-1}\bigr)+g_{-1}\bigl(q^{-2}\bigr)f_{i+1}^{(i-1)}.
\end{equation}
Using this, we check the condition for $k=1$
\[h_{i+1}^{(i)}=q^ig_1^{(0)}f_i\bigl(q^i\bigr)+g_{-1}\bigl(q^{-2}\bigr)f_{i+2}^{(i)}= q^ig_{-1}^{(0)}f_i\bigl(q^i\bigr)+\bigl(-g_1\bigl(q^2\bigr)\bigr)\cdot \bigl(-f_{i-2}^{(i)}\bigr)=h_{i-1}^{(i)}.\]

This finishes the proof of inclusion.

{\bf Conditions (1)--(3) imply being inside $\mc{A}$.}
First we prove that any element $a=\sum_{i=-N}^N w^i f_i$ such that $f_i$ have no poles and satisfy $f_{-i}\bigl(v^{-1}\bigr)=(-1)^i f_i(v)$ belongs to $\mc{A}$. Consider the elements{\samepage
\[
S_b=q^{\tfrac b2}H_b=w\frac{v^b}{v-v^{-1}}+w^{-1}\frac{v^{-b}}{v-v^{-1}} \qquad \text{and}\qquad T_b=S_{b+1}-S_{b-1}=wv^b-w^{-1}v^{-b}.
\] Then for any $l\geq 1$, we have \[T_0^{l-1}T_b=w^lv^b+(-1)^l w^{-l}v^{-b}+\sum_{i=1-l}^{l-1}w^if_i.\] Now we can prove that $a$ belongs to $\mc{A}$ using the induction on $N$.}

Now let $a=\sum_{i=-N}^N w^i f_i$ be an element of $W_0$ such that $f_i$ has simple poles satisfying the conditions~(1)--(3) above. We use induction on $N$, the base case $N=0$ being clear. We note that it is enough to find an element of $\mc{A}$ that cancels the poles of~$f_N$, as then the poles of~$f_{-N}$ will be also canceled by symmetry. We claim that \smash{$f_N^{(j,\pm)}$} is zero when $j>N$ or $j<0$. Indeed, for~${j>N}$ we use the third condition for~${i=j}$, $k=j-N$ to get
\[
f^{(j)}_{2j-N}=(-1)^{j-N+1}f^{(j)}_N.
\]
 Since~${2j-N>N}$, we have $f_{2j-N}=0$, hence the left-hand side is zero. For $j<0$, we use the third condition for~${i=j}$, $k=N-j$ to get
 \[
 f_N^{(j)}=(-1)^{N-j+1}f_{2j-N}^{(j)}=0
 \]
 since $2j-N<-N$.

We also note that \smash{$f_N^{(N,\pm)}=0$} by the second condition.

Hence, the number \smash{$f_N^{(j,\pm)}$} can be nonzero only if $0\leq j< N$. We use equation~\eqref{EqHax} for the element \smash{$x=\sum_{i=1-N}^{N-1} w^if_i$} and apply the induction hypothesis for $x$ and $N-1$. This means that for any sequence of numbers $c_0^{\pm},\ldots,c_{N-2}^{\pm}$, we can find~$x\in \mc{A}$ such that
\[
f_{N-1}^{(j,\pm)}=c_j^{\pm}.
\] We~want to prove the same for \smash{$h_N^{(0,\pm)},\ldots, h_N^{(N-1,\pm)}$}. As above, we assume that $\pm=+$, and the choices for~$\pm=-$ are done independently.

Setting $i=N$, $j<N-1$ in equation~\eqref{EqHij}, we get
\[h_N^{(j)}=g_1\bigl(q^{j+1-N}\bigr)f_{N-1}^{(j)}=g_1\bigl(q^{j+1-N}\bigr)c_j.\]

Setting $i=N$ in equation~\eqref{EqHii1} we get
\smash{$ h_N^{(N-1)}=q^{N-1}g_1^{(0)}f_{N-1}\bigl(q^{N-1}\bigr)$}. Using induction hypothesis again, we get that if we fix $c_0,\ldots,c_{N-2}$, the number $f_{N-1}\bigl(q^{N-1}\bigr)$ can be anything. Since \smash{$g_1^{(0)}$} is nonzero, we get that \smash{$h_N^{(0)}, \ldots, h_N^{(N-1)}$} can be any sequence of numbers. This finishes the proof of the statement.
\end{proof}
\section{Twisted traces}
\label{SecTwistedTraces}
The algebra $\mc{A}=\mc{A}_0\oplus\mc{A}_1$ is $\ZZ/2$-graded: the even piece $\mc{A}_{0}$ consists of elements $a=\sum_{i\in 2\ZZ} w^i f_i$, the odd piece $\mc{A}_{1}$ consists of elements $a=\sum_{i\in 2\ZZ+1} w^i f_i$. We will use this grading below.

For the next proposition, note the following. Let $a=\sum w^i f_i$ be an element of $\mc{A}$. Using~\ref{PropAInsideW}, we get $f_0(v)=f_0\bigl(v^{-1}\bigr)$, \smash{$f_0^{(0,\pm)}=0$}, and there are no other restrictions on the element $f_0$. Hence, we have a surjective linear map $a\mapsto f_0$ from $\mc{A}$ to the vector space \[V=\CC\big[v+v^{-1}\big]\oplus\bigoplus_{k\neq 0}\CC\frac{1}{\bigl(v-q^k\bigr)\bigl(v^{-1}-q^{k}\bigr)}\oplus\bigoplus_{k\neq 0}\CC\frac{1}{\bigl(v+q^k\bigr)\bigl(v^{-1}+q^{k}\bigr)}.\]
\begin{Proposition}
\label{PropTraceOnEvenElements}
 Let $g$ be an automorphism of $\mc{A}$ of the form $H_a\mapsto H_{a+k}$, $T$ be a $g$-twisted trace. Then
 \begin{itemize}\itemsep=0pt
 \item
 For an element $a\in\mc{A}_{1}$, the value $T(a)$ is uniquely defined by $T\bigl(w-w^{-1}\bigr)$ and $T\bigl(wv-w^{-1}v^{-1}\bigr)$. If $a=wf_1+w^{-1}f_{-1}$ then $T(a)=C_+f_1(\sqrt{q})+C_-f_1(-\sqrt{q})$ for some numbers~$C_+$,~$C_-$.
 \item
 There exists a linear map $S\colon V\to \CC$ such for any $a\in \mc{A}_0$, we have $T(a)=S(f_0)$.
 \end{itemize}
\end{Proposition}
\begin{proof}
 Recall that $g$ is restricted from the following automorphism of $W$, which we will also denote by $g$: $g(v)=v$, \smash{$g(w)=q^{\frac{k}{2}}v^kw=q^{-\frac{k}{2}}wv^k$}.

 We check the trace condition for a symmetric Laurent polynomial $h(v)$, it is preserved by $g$. Then for any $a=\sum_{i=-N}^N w^i f_i\in\mc{A}$, we have
$T(h(v)a)=T(ah(v))$. We have $ah(v)=\sum w^i f_ih$ and $h(v)a=\sum h(v)w^i f_i=\sum w^i h\bigl(q^{-i}v\bigr)f_i$. It follows that for any $a\in\mc{A}$ and symmetric Laurent polynomial $g$, we have
 \[T\Big(\sum w^i f_i\bigl(h(v)-h\bigl(q^{-i}v\bigr)\bigr)\Big)=0.\] Note that this element has no $w^0$ term.

 It follows from Proposition~\ref{PropAInsideW} that for any rational function $f_N$ with simple poles at $\pm 1,\ldots,\pm q^{N-1}$ we can find $f_{N-2},\ldots, f_{-N}$ such that $a=\sum w^i f_i$ belongs to $\mc{A}$. Note that $h(v)-h\bigl(q^{-N}v\bigr)$ is zero at \smash{$v=\pm q^{\frac{N}{2}}$} and this is the only condition: for $h=v+v^{-1}$, we have
 \[
 h(v)-h\bigl(q^{-N}v\bigr)=\bigl(1-q^{-N}\bigr)v+\bigl(1-q^N\bigr)v^{-1}=\bigl(1-q^{-N}\bigr)\bigl(v-q^Nv^{-1}\bigr).
\]
 Hence, for even $N$, the product $f_N\bigl(h(v)-h\bigl(q^{-N}v\bigr)\bigr)$ can be any rational function with simple poles at \[
 \pm 1,\ldots,\widehat{\pm q^{\frac{N}{2}}},\ldots,\pm q^{N-1},\]
 and for odd $N$ the product $f_N\bigl(h(v)-h\bigl(q^Nv\bigr)\bigr)$ can be any rational function vanishing at $\pm q^{\frac{N}{2}}$ with simple poles at $\pm 1,\ldots,\pm q^{N-1}$.

 By induction, it follows that for an element $a=\sum_i w^{2i+1}f_{2i+1}(v)$ we have \[T(a)=\sum_{i\geq 0,\eps=\pm}c_{2i+1,\eps}f_{2i+1}\bigl(\eps q^{\frac{2i+1}{2}}\bigr).\]

 Now we use the trace condition for $a=w-w^{-1}$ and $b=w^{2M}+w^{-2M}$ or $w^{2M}v+w^{-2M}v^{-1}$. We have $g(a)=q^{-\frac{m}{2}}\bigl(wv^m-w^{-1}v^{-m}\bigr)$. Hence,
 \begin{gather*}
 ab=w^{2M+1}-w^{2M-1}+w^{1-2M}-w^{-1-2M},\\
 bg(a)=q^{-\frac{m}{2}}\bigl(w^{2M+1}v^m-w^{2M-1}v^{-m}+w^{1-2M}v^m-w^{-2M-1}v^{-m}\bigr),
 \end{gather*}
 and a similar equation for $b=w^{2M}v+w^{-2M}v^{-1}$. Using $T(ab)=T(bg(a))$, we can express~$c_{2M+1,\pm}$ using $c_{2M-1,\pm}$. Hence, for an element \smash{$a=\sum_i w^{2i+1}f_{2i+1}(v)$}, the value $T(a)$ is uniquely defined by $T\bigl(w-w^{-1}\bigr)$ and $T\bigl(wv-w^{-1}v^{-1}\bigr)$.

 Let $N=2M$ be even. We have just showed that for any \smash{$x=\sum_{i=-M}^M w^{2i} p_{2i}(v)\in\mc{A}$} such that $p_N$ has no poles at $\pm q^{M}$, there exists \smash{$y=\sum_{i=1-N}^{N-1} w^i q_i(v)$} such that $T(x)=T(y)$ for all $g$-twisted traces $T$. Moreover, $p_0=h_0$.

 Let $a=\sum_{i=-M}^M w^{2i} p_{2i}(v)\in\mc{A}$ be any element. By induction, it follows that $T(a)$ is uniquely defined by the values of $T$ on $\CC\big[v+v^{-1}\big]$ and the values of $T$ on \[w^{2k}\frac{1}{v+q^k}+\frac{(-1)^{k+1}}{v+q^k}+\frac{(-1)^{k+1}}{v^{-1}+q^k}+w^{-2k}\frac{1}{v^{-1}+q^k}\]and
 \[w^{2k}\frac{1}{v-q^k}+\frac{(-1)^{k+1}}{v-q^k}+\frac{(-1)^{k+1}}{v^{-1}-q^k}+w^{-2k}\frac{1}{v^{-1}-q^k}.\] We see that the constant terms of these element form a basis of $V$. Define $S(f(v))=T(f(v))$ for $f\in \CC\big[v+v^{-1}\big]$ and
 \[S\left(\frac{1}{v\pm q^k}+\frac{1}{v^{-1}\pm q^k}\right)=T\left(w^{2k}\frac{1}{v\pm q^k}+\frac{(-1)^{k+1}}{v\pm q^k}+\frac{(-1)^{k+1}}{v^{-1}\pm q^k}+w^{-2k}\frac{1}{v^{-1}\pm q^k}\right).\] The proposition follows.
\end{proof}

\begin{Proposition}
\label{PropSIsAnINtegral}
 Let $T\colon\mc{A}\to \CC$ be a $g$-twisted trace and $S\colon V\to \CC$ be the function from Proposition~{\rm\ref{PropTraceOnEvenElements}}. Then
 \[S(F(z))=\int_{\eps S^1}F(z)\omega(z)\frac{{\rm d}z}{z},\] where $S^1$ is the unit circle, $\omega(z)$ is a holomorphic function on $\CC^{\times}$ such that $\omega(qz)=q^{-\frac{m}{2}}z^{-m}\omega(z)$ and $\omega(z)=\omega\bigl(z^{-1}\bigr)$, and $\eps$ is any number smaller than all of the absolute values of the poles of~$F$. Conversely, every such $S$ defines a trace by the rule $T\bigl(\sum_i w^i f_i\bigr)=S(f_0)$. In particular, the dimension of the space of traces $T$ that vanish on $\mc{A}_1$ is $1+\frac{m}{2}$.
\end{Proposition}
\begin{proof}
 The proof consists of three steps. First, we rewrite the trace condition for $T|_{\mc{A}_0}$ in terms of $S$ and get equation~\eqref{EqSGeneralCondition}. Then we check that each $S$ as in the statement of the proposition satisfies~\eqref{EqSGeneralCondition} and show that the space of possible functions $\omega(z)$ has dimension $\frac{m}{2}+1$. Finally, we show that the dimension of the space of $S\colon V\to \CC$ satisfying~\eqref{EqSGeneralCondition} has dimension at most~${\frac{m}{2}+1}$.

 We use the trace condition $T(ba)=T(ag(b))$ for
 \[
 a=\sum w^{2i+1}f_{2i+1} \qquad \text{and} \qquad b=w\smash{\frac{v^k}{v-v^{-1}}}+w^{-1}\smash{\frac{v^{-k}}{v-v^{-1}}}.
 \]
 Since $ba$ and $ag(b)$ belong to $\mc{A}_0$, by Proposition~\ref{PropTraceOnEvenElements}, we have
 \[
 T(ba)=S\bigl(\big[w^0\big](ba)\bigr) \qquad \text{and} \qquad T(ag(b))=S\bigl(\big[w^0\big]\bigl(ag(b)\bigr)\bigr).
 \]
 We note that the only condition on $f_1(v)$ is that it is regular at $v=\pm q$. We have
 \[
 f_{-1}(v)=-f_1\bigl(v^{-1}\bigr).
 \]
  Denote $f_1(v)=F(v)$, so that $f_{-1}(v)=-F\bigl(v^{-1}\bigr)$. We~have
 \begin{align*}
 \big[w^0\big](ba)&=w\frac{v^k}{v-v^{-1}}\cdot\bigl(-w^{-1}F\bigl(v^{-1}\bigr)\bigr)+w^{-1}\frac{v^{-k}}{v-v^{-1}}wF(v)\\
 &=q^k\bigg(\frac{-v^k}{qv-q^{-1}v^{-1}}F\bigl(v^{-1}\bigr)+\frac{v^{-k}}{q^{-1}v-qv^{-1}}F(v)\bigg).
 \end{align*}
 Note that $g(b)=q^{-\frac{m}2}\bigl(w\frac{v^{k+m}}{v-v^{-1}}+w^{-1}\frac{v^{-k-m}}{v-v^{-1}}\bigr)$. Hence,
 \begin{align*}
 \big[w^0\big](ag(b))&=q^{-\frac{m}{2}}\bigg(wF(v)w^{-1}\frac{v^{-k-m}}{v-v^{-1}}-w^{-1}F\bigl(v^{-1}\bigr)w\frac{v^{k+m}}{v-v^{-1}}\bigg)\\
 &=q^{-\frac{m}2}\bigg(F(qv)\frac{v^{-k-m}}{v-v^{-1}}-F\bigl(qv^{-1}\bigr)\frac{v^{k+m}}{v-v^{-1}}\bigg).
 \end{align*}

 The trace condition for $T$ implies
 \begin{gather}
 q^kS\bigg(\frac{-v^k}{qv-q^{-1}v^{-1}}F\bigl(v^{-1}\bigr)+\frac{v^{-k}}{q^{-1}v-qv^{-1}}F(v)\bigg)\nonumber\\
 \qquad=q^{-\frac{m}2}S\bigg(F(qv)\frac{v^{-k-m}}{v-v^{-1}}-F\bigl(qv^{-1}\bigr)\frac{v^{k+m}}{v-v^{-1}}\bigg).\label{EqSGeneralCondition}
 \end{gather}

 Conversely, we claim that every $S$ satisfying~\eqref{EqSGeneralCondition} gives a trace $T$ by $T\bigl(\sum w^i f_i\bigr)=S(f_0)$. The algebra $\mc{A}$ is generated by $H_0$, $H_1$ and $v+v^{-1}$. Hence, it is enough check the trace condition~${T(ba)=T(ag(b))}$ for $b=H_0,H_1$ or $b=v+v^{-1}$. In the case when $b=v+v^{-1}$, the trace condition is satisfied for any $S$. In the case when $b=H_k$, it is enough to take $a\in \mc{A}_1$. In this case, the computation above shows that $T(ba)=T(ag(b))$ is equivalent to~\eqref{EqSGeneralCondition}.

 Now we check that the integral from the statement of the proposition satisfies the equation~\eqref{EqSGeneralCondition}
 \begin{gather*}
 q^k \left(\int_{\eps S^1} \left(\frac{-z^k}{qz-q^{-1}z^{-1}}F\bigl(z^{-1}\bigr)+\frac{z^{-k}}{q^{-1}z-qz^{-1}}F(z)\right)\omega(z)\frac{{\rm d}z}{z}\right)\\
\qquad =q^{-\frac{m}2}\left(\int_{\eps S^1}\left(F(qz)\frac{z^{-k-m}}{z-z^{-1}}-F\bigl(qz^{-1}\bigr)\frac{z^{k+m}}{z-z^{-1}}\right)\omega(z)\frac{{\rm d}z}{z}\right)
 .
 \end{gather*}
 Indeed,
 \begin{gather*}
 \int_{\eps S^1}\bigg(F(qz)\frac{z^{-k-m}}{z-z^{-1}}-F\bigl(qz^{-1}\bigr)\frac{z^{k+m}}{z-z^{-1}}\bigg)\omega(z)\frac{{\rm d}z}{z}\\
\qquad= \int\limits_{q^{-1}\eps S^1}F(qz)\frac{z^{-k-m}}{z-z^{-1}}\omega(z)\frac{{\rm d}z}{z}-\int\limits_{q\eps S^1}F\bigl(qz^{-1}\bigr)\frac{z^{k+m}}{z-z^{-1}}\omega(z)\frac{{\rm d}z}{z}\\
 \qquad= \int_{\eps S^1}F(z)\frac{q^{k+m}z^{-k-m}}{q^{-1}z-qz^{-1}}\omega(q^{-1}z)\frac{{\rm d}z}{z}-\int_{\eps S^1}F\bigl(z^{-1}\bigr)\frac{q^{k+m}z^{k+m}}{qz-q^{-1}z^{-1}}\omega(qz)\frac{{\rm d}z}{z}\\
 \qquad= q^{k+\frac{m}{2}}\bigg(\int_{\eps S^1}F(z)\frac{z^{-k}}{q^{-1}z-qz^{-1}}\omega(z)\frac{{\rm d}z}{z}-\int_{\eps S^1}F\bigl(z^{-1}\bigr)\frac{z^k}{qz-q^{-1}z^{-1}}\omega(z)\frac{{\rm d}z}{z}\bigg).
 \end{gather*}
 We shifted the contours, changed the variable and used that $\omega(qz)=z^{-m}q^{-\frac{m}{2}}\omega(z)$, so that $\omega(q^{-1}z)=z^m q^{-\frac{m}{2}}\omega(z)$.

 This shows that every function $\omega$ satisfying
 \[
 \omega(qz)=q^{-\frac{m}{2}}z^{-m}\omega(z) , \qquad \omega(z)=\omega\bigl(z^{-1}\bigr)
 \]
gives a~trace. The space of such functions has dimension $1+\frac{m}{2}$. To show that, consider \smash{$w(z)=\omega\bigl({\rm e}^{2\pi {\rm i} z}\bigr)$}. Let \smash{$q={\rm e}^{2\pi {\rm i} \tau}$}. Then
\[
w(z+1)=w(z) , \qquad w(-z)=w(z) \qquad \text{and} \qquad w(z+\tau)={\rm e}^{-\pi \mathrm{i} m(\tau+2z)}w(z).
\]
 Any such function can be obtained as \smash{$C\prod_{j=1}^{\frac m2} \vartheta(z-\alpha_j)\vartheta(z+\alpha_j)$}, where $\vartheta$ is Jacobi theta function and~$\alpha_j$ and $C$ are any numbers. 

 Now we prove that all maps $S$ satisfying~\eqref{EqSGeneralCondition} can be obtained in this way. It is enough to prove that the dimension of the space of possible $S$ is at most $\frac{m}{2}+1$. Since any $S$ satisfying~\eqref{EqSGeneralCondition} defines a trace $T$, we will use the trace condition below.

 Taking regular $F$ such that $F(q)=0$ and $F(-q)\neq 0$, we see that the argument of $S$ on the left-hand side of~\eqref{EqSGeneralCondition} has a pole at $v=-q^{\pm 1}$ and is regular at $v=q^{\pm 1}$. The argument on~the right-hand side of~\eqref{EqSGeneralCondition} is regular. Hence, \smash{$S\bigl(\frac{1}{(v+ q)(v^{-1}+q)}\bigr)$} is defined by the restriction of~$S$~to $\CC\big[v+v^{-1}\big]$. Taking regular $F$ such that $F(-q)=0$ and $F(q)\neq 0$, we get that \smash{$S\bigl(\frac{1}{(v-q)(v-q^{-1})}\bigr)$}~is defined by the restriction of~$S$~to $\CC\big[v+v^{-1}\big]$. Now, suppose that~$F$ has a~pole at $q^k$ for some~$k\neq 1$. Then the argument of~$S$~on the left-hand side of~\eqref{EqSGeneralCondition} has a pole at~$\pm q^{\pm 1}$ and~$q^{\pm k}$ (no pole if~${k=0}$) and the right-hand side has a pole at $q^{\pm(k-1)}$. It follows by induction with the base case $k=1$ that \smash{$S\bigl(\frac{1}{(v- q^k)(v^{-1}- q^k)}\bigr)$} is defined by the restriction of $S$ to $\CC\big[v+v^{-1}\big]$. A~similar~argument shows that \smash{$S\bigl(\frac{1}{(v+q^k)(v^{-1}+q^k)}\bigr)$} is defined by the restriction of $S$ to $\CC\big[v+v^{-1}\big]$.

 Take the same $a\in\mc{A}$ as before and $b=-w+w^{-1}$. Writing $T(ba)=T(ag(b))$ in terms of $S$, we get
 \[S\bigl(F(v)+F\bigl(v^{-1}\bigr)\bigr)=q^{-\tfrac m2}S\bigl(F(qv)v^{-m}+F\bigl(qv^{-1}\bigr)v^m\bigr).\]
 For $F(v)=v^k$, we get
\[
S\bigl(v^k+v^{-k}\bigr)=q^{k-\tfrac m2}S\bigl(v^{k-m}+v^{m-k}\bigr).
\]
 It follows that $S$ is uniquely defined by $S(1), S\bigl(v+v^{-1}\bigr),\dots, S\bigl(v^{\frac{m}{2}}+v^{-\frac{m}{2}}\bigr)$. Therefore, the dimension of the space of linear maps $S$ satisfying~\eqref{EqSGeneralCondition} is at most $1+\frac{m}{2}$, as claimed.
\end{proof}

\section{Positive traces}
\label{SecPositiveTraces}
\begin{Proposition}
\label{PropWHasZeroesAtPlusMinusOne}
 Let $T$ be a positive trace. Let $\omega$ be a function from Proposition~{\rm\ref{PropSIsAnINtegral}}. Then $\omega(1)=\omega(-1)=0$.
\end{Proposition}
\begin{proof}
 Assume the opposite. Then we can write
 \[S(F(z))=\int_{S^1}F(z)\omega(z)\frac{{\rm d}z}{z}+\sum_{i>0} c_{i,\pm}\operatorname{Res}_{\pm q^i}F,\] where $c_{i,\pm}$ are nonzero multiples of $\omega(\pm 1)$, so that for at least one of the choices of sign all $c_{i,\pm}$ are nonzero. Without loss of generality, assume that $c_i=c_{i,+}$ are nonzero.

 Let $a=\bigl(w^2F(v)-w^{-2}vF\bigl(v^{-1}\bigr)\bigr)(v-1)^{-1}$, where $F(v)$ is regular. This is an element of $\mc{A}$. Then \[\rho(a)=q^m\bigl(-w^2v^{1-m}\overline{F}(v)+w^{-2}v^{m}\overline{F}\bigl(v^{-1}\bigr)\bigr)(v-1)^{-1}\] and
 \begin{align*}
 \big[w^0\big]a\rho(a)={}&q^m\bigl(w^2F(v)(v-1)^{-1}w^{-2}v^m\overline{F}\bigl(v^{-1}\bigr)(v-1)^{-1}\\
 &+w^{-2}vF\bigl(v^{-1}\bigr)(v-1)^{-1}w^2\ovl{F}(v)v^{1-m}(v-1)^{-1}\bigr).
 \end{align*}
 We compute
 \begin{gather*}
 w^2F(v)(v-1)^{-1}w^{-2}v^m\ovl{F}\bigl(v^{-1}\bigr)(v-1)^{-1}+w^{-2}vF\bigl(v^{-1}\bigr)(v-1)^{-1}w^2v^{1-m}\ovl{F}(v)(v-1)^{-1}\\
\qquad= F\bigl(q^2v\bigr)\ovl{F}\bigl(v^{-1}\bigr)\bigl(q^2v-1\bigr)^{-1}v^m(v-1)^{-1}\\
\phantom{\qquad=}{}+F\bigl(q^{2}v^{-1}\bigr)\ovl{F}(v)q^{-2}v\bigl(q^{-2}v-1\bigr)^{-1}v^{1-m}(v-1)^{-1}=G(v)+G\bigl(v^{-1}\bigr),
 \end{gather*}
 where
 \[G(v)=\ovl{F}\bigl(v^{-1}\bigr)F\bigl(q^2v\bigr)\cdot\frac{v^m}{\bigl(q^2v-1\bigr)(v-1)}\]
 and
 \[G\bigl(v^{-1}\bigr)=\ovl{F}(v)F\bigl(q^2 v^{-1}\bigr)\cdot \frac{q^{-2}v^{2-m}}{(q^{-2}v-1)(v-1)}.\]
 We have
 \begin{equation}
 \label{EqTaRhoaAsIntegral}
 T(a\rho(a))=S\bigl(\big[w^0\big]a\rho(a)\bigr)=S\bigl(G(v)+G\bigl(v^{-1}\bigr)\bigr)=\int_{\eps S^1} \bigl(G(z)+G\bigl(z^{-1}\bigr)\bigr)\omega(z)\frac{{\rm d}z}{z}.
 \end{equation}
 We now evaluate this integral.
 The only pole of $G$ between $q^{-1}S^1$ and $0$ is $v=1$, and the residue~is
 \[\ovl{F}(1)\frac{F\bigl(q^2\bigr)}{q^2-1}.\]
 The only pole of $G\bigl(z^{-1}\bigr)$ between $0$ and $qS^1$ is $v=q^2$, and the residue is
\[\ovl{F}(q^{2})\cdot\frac{F(1)q^{-2}q^{2(2-m)}}{\bigl(q^2-1\bigr)q^{-2}}=F(1)\ovl{F}(q^{2})\frac{q^{2(2-m)}}{q^2-1}.\]
Therefore,
 \begin{align}
\int_{\eps S^1} \bigl(G(z)+G\bigl(z^{-1}\bigr)\bigr)\omega(z)\frac{{\rm d}z}{z}=&c_0\ovl{F}(1)\frac{F\bigl(q^2\bigr)}{q^2-1}+c_1F(1)\ovl{F}\bigl(q^{2}\bigr)\frac{q^{2(2-m)}}{q^2-1}\nonumber\\
&+\int_{q^{-1}S^1}G(z)\omega(z)\frac{{\rm d}z}{z}+\int_{qS^1}G\bigl(z^{-1}\bigr)\omega(z)\frac{{\rm d}z}{z} . \label{EqIntegralAsResiduePlusSymmetricG}
\end{align}
We rewrite the integrals
 \begin{align*}
 \int_{q^{-1}S^1}G(z)\omega(z)\frac{{\rm d}z}{z}&=\int_{S^1}G\bigl(q^{-1}z\bigr)\omega\bigl(q^{-1}z\bigr)\frac{{\rm d}z}{z}\\
 &=\int_{S^1}\ovl{F}\bigl(qz^{-1}\bigr)F(qz)\frac{q^{-m}v^m}{(qv-1)\bigl(q^{-1}v-1\bigr)}\omega\bigl(q^{-1}z\bigr)\frac{{\rm d}z}{z},
\\
 \int_{qS^1}G\bigl(z^{-1}\bigr)\omega(z)\frac{{\rm d}z}{z}&=\int_{S^1}G\bigl(q^{-1}z^{-1}\bigr)\omega(qz)\frac{{\rm d}z}{z}\\
 &=\int_{S^1}\ovl{F}(qz)F\bigl(q z^{-1}\bigr)\cdot \frac{q^{-m}v^{z-m}}{(v-1)\bigl(q^2v-1\bigr)}\cdot \omega(qz)\frac{{\rm d}z}{z}.
 \end{align*}
Now take $F(z)=q^{-N}z^N-a$ for some constant $a$ and natural number $N$. For large $N$, the leading term of
\[c_0\ovl{F}(1)\frac{F\bigl(q^2\bigr)}{q^2-1}+c_1F(1)\ovl{F}(q^{2})\frac{q^{2(2-m)}}{q^2-1}\] is $C_0aq^{-N}+C_1\ovl{a}q^{-N}$, where $C_0$, $C_1$ are constant numbers. For some choice of $a$, this is a negative number with absolute value growing as $q^{-N}$.

On the other hand, for $z\in S^1$ polynomials $\ovl{F}\bigl(qz^{-1}\bigr)$, $F(qz)$, $\ovl{F}(qz)$, $F\bigl(qz^{-1}\bigr)$ have absolute value at most $1+\lvert a\rvert$. Hence, the integrals in the right-hand side of~\eqref{EqIntegralAsResiduePlusSymmetricG} are bounded when $N$ tends to infinity. On the other hand, the first two terms in the right-hand side of~\eqref{EqIntegralAsResiduePlusSymmetricG} tend to minus infinity. Comparing~\eqref{EqTaRhoaAsIntegral} and~\eqref{EqIntegralAsResiduePlusSymmetricG}, we see that for large enough $N$ we have $T(a\rho(a))<0$, a contradiction with positivity.
 \end{proof}

 In particular, for $m=4$ we get that the function $\omega$ is unique up to a constant. Indeed, consider the function $w$ such that $w(z)=\omega\bigl({\rm e}^{2\pi {\rm i} z}\bigr)$. Then, using $w(0)=w(\tfrac12)=0$, we get $w(z)=\vartheta\bigl(z-\frac{\tau}{2}\bigr)\vartheta\bigl(z+\frac{\tau}{2}\bigr)\vartheta\bigl(z-\tfrac12-\frac{\tau}{2}\bigr)\vartheta\bigl(z+\frac{1}{2}+\frac{\tau}{2}\bigr)$ up to a constant. Here $\vartheta(z)$ is the Jacobi theta function.
\begin{Proposition}
\label{PropPositiveTraceIsZeroOnOddElements}
 Let $T$ be a positive trace. Then $T$ vanishes on $\mathcal{A}_1$.
\end{Proposition}
 \begin{proof}
 Using Proposition~\ref{PropTraceOnEvenElements}, it is enough to prove that $T(a)=0$ for elements $a$ of the form $a=wF(v)-wF\bigl(v^{-1}\bigr)$. For these elements, we have $T(a)=C_+F(\sqrt{q})+C_-F(-\sqrt{q})$. Assume that $C_+$ is nonzero.

 Let $a=wF(v)-w^{-1}F\bigl(v^{-1}\bigr)$ for some Laurent polynomial $F$, $b=a+1$. Then \[\rho(b)=q^{-\frac{m}{4}}\bigl(-wv^{\frac{m}{2}}\ovl{F}(v)+w^{-1}v^{-\frac{m}{2}}\ovl{F}\bigl(v^{-1}\bigr)\bigr)+1.\]
 Note that $T(b\rho(b))$ depends only on the $w^0$ and $w^1$ coefficients of $b\rho(b)$. They can be computed as follows
 \begin{gather*}
 \big[w^0\big]b\rho(b)=q^{-\frac{m}{4}}\bigl(F(qv)v^{-\frac{m}{2}}\ovl{F}\bigl(v^{-1}\bigr)+F\bigl(qv^{-1}\bigr)v^{\frac{m}{2}}\ovl{F}(v)\bigr)+1,\\
 \big[w^1\big]b\rho(b)=F(v)-v^{\frac{m}{2}}\ovl{F}(v).
 \end{gather*}
 Let $G=F(\sqrt{q}v)$, $I(G)=\int_{S^1}\lvert G(z)\rvert^2 \frac{{\rm d}z}{z}$. We have $S\bigl(\big[w^0\big]b\rho(b)\bigr)\leq cI(G)$ for some $c>0$. On the other hand, trace for the odd powers of $w$ is a linear combination of $G(1)$, $G(-1)$, $\ovl{G}(1)$ and~$\ovl{G}(-1)$; and $G(1)$ has nonzero coefficient. Let \smash{$G_N(z)=a\bigl(\frac{z+1}{2}\bigr)^N$}. Then $G_N(1)=a$, $G_N(-1)=0$. On the other hand, $G_N(z)$ is a sequence of functions such that $\lvert G_N(z)\rvert\leq 1$ tends to zero for almost all $z$ as $N$ tends to infinity. Since pointwise convergence implies convergence in measure for spaces of finite total measure and the functions $G_N$ are uniformly bounded, we deduce that $I(G_N(z))$ tends to zero as $N$ tends to infinity. We get that $T(b\rho(b))$ tends to a~linear combination of $a$ and $\ovl{a}$. We can choose $a$ so that this linear combination does not belong to~$\mathbb{R}_{\geq 0}$. We get a contradiction with positivity of $T$.
 \end{proof}

 Let $T$ be a positive trace. Combining Propositions~\ref{PropWHasZeroesAtPlusMinusOne} and~\ref{PropPositiveTraceIsZeroOnOddElements}, it follows that for $a=\sum w^i f_i$ we have $T(a)=\int f_0(z)\omega(z)\frac{{\rm d}z}{z}$ and $\omega(1)=\omega(-1)=1$. 
\begin{Proposition}
\label{PropPositivityForW}
 Let $T$ be a trace on $\mc{A}$ of the form $T(a)=\int f_0(z)\omega(z)\frac{{\rm d}z}{z}$, where $w$ is a holomorphic function on $\CC^*$ such that \smash{$\omega(qz)=q^{-\frac{m}{2}}z^{-m}\omega(z)$}, $\omega(z)=\omega\bigl(z^{-1}\bigr)$ and $\omega(1)=\omega(-1)=0$. Then $T$ is positive if and only if $\omega(z)$ and \smash{$z^{-\frac{m}{2}}\omega\bigl(\sqrt{q^{-1}}z\bigr)$} are nonnegative on $S^1$.
\end{Proposition}
 \begin{proof}
 First we check that any such function $\omega$ indeed gives a positive trace.

 Let $a=\sum w^k f_k$, then \[\rho(a)=\sum q^{-\frac{k^2m}{4}}w^{-k}v^{-\frac{km}{2}}\ovl{f_k}\bigl(v^{-1}\bigr)\] and
 \begin{equation}
 \label{EqaRhoaAnyelement}
 \big[w^0\big]a\rho(a)=\sum q^{-\frac{k^2m}{4}} f_k(q^k v)v^{-\frac{km}{2}}\ovl{f_k}\bigl(v^{-1}\bigr).
 \end{equation}
 It is enough to show that \[I(f_k)=\int_{\eps S^1} f_k\bigl(q^k z\bigr)z^{-\frac{km}{2}}\ovl{f_k}\bigl(z^{-1}\bigr)\omega(z)\frac{{\rm d}z}{z}\] is positive for any $k$ such that $f_k$ is nonzero. Let $f=f_k$, this is a function with the poles of order at most one at $\pm q^k$, $k\in \ZZ$.
 Then the only possible poles of $f\bigl(q^k z\bigr)\ovl{f}\bigl(z^{-1}\bigr)$ are poles of order at most two at $\pm q^k$, $k\in \ZZ$. Since $\omega(1)=\omega(-1)=0$ and $\omega(z)=\omega\bigl(z^{-1}\bigr)$, the function~$\omega$ has a double root at $1$ and $-1$. Using the quasi-periodicity condition, we get that $w$ has a~double root at $\pm q^k$, $k\in \ZZ$. It follows that $f\bigl(q^k z\bigr)\ovl{f}\bigl(z^{-1}\bigr)\omega(z)$ is a holomorphic function on $\CC^*$. Hence, we can write
 \begin{align*}
 I(f)&=\int_{\eps S^1} f\bigl(q^k z\bigr)z^{-\frac{km}{2}}\ovl{f}\bigl(z^{-1}\bigr)\omega(z)\frac{{\rm d}z}{z}=\int_{q^{-\frac{k}{2}}S^1}f\bigl(q^k z\bigr)z^{-\frac{km}{2}}\ovl{f}\bigl(z^{-1}\bigr)\omega(z)\frac{{\rm d}z}{z}\\
& = q^{\frac{k^2m}{4}}\int_{S^1}f\bigl(q^{\frac{k}{2}}t\bigr)\ovl{f}\bigl(q^{\frac{k}{2}}t^{-1}\bigr)t^{-\frac{km}{2}}\omega\bigl(q^{-\frac{k}{2}}t\bigr)\frac{{\rm d}t}{t}.
 \end{align*}

 For $z\in S^1$, we have $f\bigl(q^{\frac{k}{2}}t\bigr)\ovl{f}\bigl(q^{\frac{k}{2}}t^{-1}\bigr)=\big\lvert f\bigl(q^{\frac{k}{2}}t\bigr)\big\rvert^2$. It remains to show that $z^{-\frac{km}{2}}\omega\bigl(q^{-\frac{k}{2}}z\bigr)$ is nonnegative on $S^1$. We have
 \[\omega\bigl(q^{1-\frac{k}{2}}zz\bigr)=q^{-\frac{m}{2}}\bigl(q^{-\frac{k}{2}}z\bigr)^{-m}\omega\bigl(q^{-\frac{k}{2}}z\bigr),\] so that
 \[
 z^{\frac{(2-k)m}{2}}\omega\bigl(q^{\frac{2-k}{2}}z\bigr)=q^{\frac{m(k-1)}{2}}z^{-\frac{km}{2}}\omega\bigl(q^{-\frac{k}{2}}z\bigr),
 \] so that the sign of \smash{$z^{-\frac{km}{2}}\omega\bigl(q^{-\frac{k}{2}}z\bigr)$} depends only on the parity of $k$. For $k=0,1$, we get exactly the condition in the proposition statement.

 Now we prove the other direction of the proposition statement. Let $T$ be a positive trace. Take $a=f(v)$. Then \[T(a\rho(a))=\int_{S^1}f(z)\ovl{f}\bigl(z^{-1}\bigr)\omega(z)\frac{{\rm d}z}z\] should be greater than zero for all nonzero Laurent polynomials $f$ such that $f(z)=f\bigl(z^{-1}\bigr)$. Using the density of trigonometric polynomials in $L^2\bigl(S^1\bigr)$ and the condition $\omega(z)=\omega\bigl(z^{-1}\bigr)$, we~get that $\omega(z)$ should be nonnegative on $S^1$.

 Take $a=wF(v)-w^{-1}F\bigl(v^{-1}\bigr)\in\mc{A}$, where $F$ is a Laurent polynomial. Using~\eqref{EqaRhoaAnyelement}, we get
 \[\big[w^0\big]a\rho(a)=q^{-\frac{m}{4}}\bigl(F(qv)v^{-\frac{m}{2}}\ovl{F}\bigl(v^{-1}\bigr)+F\bigl(qv^{-1}\bigr)v^{\frac{m}{2}}\ovl{F}(v)\bigr),\] so that
 \[T(a\rho(a))=q^{-\frac{m}{4}}\int_{\eps S^1}\bigl(F(qz)z^{-\frac{m}{2}}\ovl{F}\bigl(z^{-1}\bigr)+F\bigl(qz^{-1}\bigr)z^{\frac{m}{2}}\ovl{F}(z)\bigr)\omega(z)\frac{{\rm d}z}{z}.\]
 Using $\omega(z)=\omega\bigl(z^{-1}\bigr)$, we get
 \[\int_{\eps S^1}F\bigl(qz^{-1}\bigr)z^{\frac{m}{2}}\ovl{F}(z)\omega(z)\frac{{\rm d}z}{z}=\int_{\eps^{-1}S^1}F(qz)z^{-\frac{m}{2}}\ovl{F}\bigl(z^{-1}\bigr)\omega(z)\frac{{\rm d}z}{z}.\]
 Shifting contours, we get
 \begin{align*}
 T(a\rho(a))={}&2q^{-\frac{m}{4}}\int_{\sqrt{q^{-1}}S^1}F(qz)\ovl{F}\bigl(z^{-1}\bigr)z^{-\frac{m}{2}}\omega(z)\frac{{\rm d}z}{z}\\
 ={}&2\int_{S^1}F(\sqrt{q}t)\ovl{F}\bigl(\sqrt{q}t^{-1}\bigr)t^{-\frac{m}{2}}\omega\bigl(\sqrt{q^{-1}}t\bigr)\frac{{\rm d}t}{t}.
 \end{align*}
 Using the fact that trigonometric polynomials are dense in $L^2\bigl(S^1\bigr)$, we get that $t^{-\frac{m}{2}}\omega\bigl(\sqrt{q^{-1}}t\bigr)$ is nonnegative on $S^1$. The proposition follows.
 \end{proof}

 \begin{Theorem}
 \label{ThrAnswerKTheory}
 Positive traces on $\mc{A}$ are in one-to-one correspondence with holomorphic functions $\omega$ on $\CC^{\times}$ such that
 \begin{enumerate}\itemsep=0pt
 \item[$(1)$] $\omega(qz)=q^{-\frac{m}{2}}z^{-m}\omega(z)$;
 \item[$(2)$] $\omega(z)=\omega\bigl(z^{-1}\bigr)$;
 \item[$(3)$] $\omega(1)=\omega(-1)=0$;
 \item[$(4)$] $\omega(z)$ and $z^{-\frac{m}{2}}\omega\bigl(\sqrt{q^{-1}}z\bigr)$ are nonnegative on $S^1$.
 \end{enumerate}

 The convex cone of such functions $\omega$ has real dimension $\frac{m}{2}-1$. In particular, for $m=4$ there is a unique positive trace up to a constant.
 \end{Theorem}
 \begin{proof}
 Let $T$ be a positive trace. Using Proposition~\ref{PropPositiveTraceIsZeroOnOddElements} we get that $T$ vanishes on $\mc{A}_1$. Then, by Propositions~\ref{PropTraceOnEvenElements} and~\ref{PropSIsAnINtegral},
 \[T\left(\sum_{i=-N}^N w^i f_i\right)=\int f_0(z)\omega(z)\frac{{\rm d}z}{z}\] for a function $\omega$ satisfying the first two conditions. Using Proposition~\ref{PropWHasZeroesAtPlusMinusOne}, we get $\omega(1)={\omega(-1)=0}$. For such $T$ and $\omega$, Proposition~\ref{PropPositivityForW} says that the fourth condition on $\omega$ is equivalent to $T$ being positive.

 We turn to the second statement. Let $w$ be a functions satisfying conditions (1)--(4). Consider~$w(z)$ such that $w(z)=\omega\bigl({\rm e}^{2\pi {\rm i} z}\bigr)$. Let $q={\rm e}^{2\pi {\rm i}\tau}$. Conditions~(1) and~(4) mean that $w$ gives a~positive trace on $q$-Weyl algebra. The set of the corresponding functions $w$ is described in~\cite[Theorem~3.7]{KQWeyl}. Namely, the function $w$ satisfies conditions~(1) and~(4) if and only if there exist $\lambda>0$ and $\alpha_1,\ldots, \alpha_m$ divided into pairs $\alpha_i=\ovl{\alpha_j}$ such that
 \begin{equation}
 \label{EqWProductOfVarThetas}
 w(z)=\lambda\prod_{i=1}^m \vartheta(z-\alpha_i).
 \end{equation} Here $\vartheta$ is the Jacobi theta function. It follows from \cite[Lemma 3.6]{KQWeyl} that we can assume $\lvert\Im\alpha_i\rvert<\Im\tau$. Note that $\tau$ in this article is the same as $\tau$ in~\cite{KQWeyl}, but our $q$ corresponds to $q^2$ in~\cite{KQWeyl}.

Condition~(2) is equivalent to $w(z)=w(-z)$. Evaluating at $z=\tfrac12+\tfrac{\tau}{2}-\alpha_k$, we get
\[
\prod_{j=1}^m \vartheta\bigl(\tfrac12+\tfrac{\tau}{2}-\alpha_k-\alpha_j\bigr)=0.
\]
 Hence, from any $k$ from $1$ to $m$, there exists $j$ such that $\alpha_j+\alpha_k\in \ZZ+\ZZ\tau$.

We claim that we can change $\lambda,\alpha_1,\ldots,\alpha_m$ so that~\eqref{EqWProductOfVarThetas} is still satisfied and $\alpha_1,\ldots,\alpha_m$ are divided into quadruples $\alpha_j$, $-\alpha_j$, $\ovl{\alpha_j}$, $-\ovl{\alpha_j}$ and pairs $\alpha_k$, $\ovl{\alpha_k}$ with $\Re\alpha_k\in\tfrac12\ZZ$.

 Let $j$, $k$ satisfy $\alpha_j+\alpha_k\in \ZZ+\ZZ\tau$. There are three cases:
 \begin{enumerate}\itemsep=0pt
 \item[$(1)$]
Assume that $\alpha_j\neq \alpha_k$ or $\ovl{\alpha_k}$. Shifting $\alpha_j$ by an integer we get $\alpha_j+\alpha_k\in\ZZ\tau$. Since $\lvert\Im\alpha_{j,k}\rvert<\Im\tau$ we have $\alpha_j+\alpha_k\in\{-\tau,0,\tau\}$.

Assume that $\alpha_j+\alpha_k=-\tau$. Note that \smash{$\vartheta(z-\alpha_j-\tau)={\rm e}^{\pi {\rm i}(-\tau+2z-2\alpha_j)}\vartheta(z-\alpha_j)$} and \smash{$\vartheta(z-\ovl{\alpha_j}+\tau)={\rm e}^{-\pi {\rm i}(\tau+2z-2\ovl{\alpha_j})}\vartheta(z-\ovl{\alpha_j})$}, so that
\[\vartheta(z-\alpha_j-\tau)\vartheta(z-\ovl{\alpha_j}+\tau)={\rm e}^{-2\pi {\rm i} (\tau+\alpha_j-\ovl{\alpha_j})}\vartheta(z-\alpha_j)\vartheta(z-\ovl{\alpha_j}).\] Hence, we can change $\alpha_j$ and $\ovl{\alpha_j}$ to $\alpha_j+\tau$ and $\ovl{\alpha_j+\tau}=\ovl{\alpha_j}-\tau$. This will multiply $\lambda$ by a~positive number.
 After this change, we have $\alpha_j+\alpha_k=0$. Similarly, if $\alpha_j+\alpha_k=\tau$, we can shift $\alpha_j$ to get $\alpha_j+\alpha_k=0$.
Hence, in the case when $\alpha_j+\alpha_k\in \ZZ+\ZZ\tau$ and $\alpha_j\neq \alpha_k,\ovl{\alpha_k}$, we get a quadruple $\alpha_k$, $\ovl{\alpha_k}$, $-\alpha_k$, $-\ovl{\alpha_k}$.
 \item[$(2)$]
 If $\alpha_j=\alpha_k$ we get $\alpha_k\in \tfrac{1}{2}\ZZ+\frac{\tau}{2}\ZZ$. Hence, $\alpha_k+\ovl{\alpha_k}\in \ZZ$.
 \item[$(3)$]
 If $\alpha_j=\ovl{\alpha_k}$, we get $\alpha_k+\ovl{\alpha_k}\in \ZZ+\ZZ\tau$. Since $\alpha_k+\ovl{\alpha_k}$ is a real number, this means $\alpha_k+\ovl{\alpha_k}\in \ZZ$.
 \end{enumerate}

We get that $\alpha_1,\dots,\alpha_m$ are divided in quadruples $\alpha_j$, $-\alpha_j$, $\ovl{\alpha_j}$, $-\ovl{\alpha_j}$ and pairs $\alpha_k$, $\ovl{\alpha_k}$ with~$\Re\alpha_k\in \tfrac{1}{2}\ZZ$. Conversely, for such sequence $\alpha_j$ the function $\omega(z)=\lambda \prod_{i=1}^m \vartheta(z-\alpha_i)$ satisfies $w(z)=w(-z)$.

The condition (3) just means that two of the pairs are $\pm\frac{\tau}{2}$ and $\frac{1}{2}\pm\frac{\tau}{2}$.

After that, each quadruple has one complex degree of freedom and each tuple has one real degree of freedom, giving real dimension $\frac{m-4}{2}=\frac{m}{2}-2$. A choice of $\lambda$ is an additional degree of freedom. So we get a cone of real dimension $\frac{m}{2}-1$, as claimed.
 \end{proof}

\subsection*{Acknowledgements}
The first part of the article was written by J.V.\ under the mentorship of D.K.\ at Research Science Institute, hosted by the Massachusetts Institute of Technology. We would like to thank Pavel Etingof for suggesting this project and for comments on the earlier version of the paper.
We would also like to thank RSI Head Mentor, Dr.~Tanya Khovanova for comments on the earlier versions of the paper.
J.V.\ also thanks Peter Gaydarov
for helpful suggestions in preparing this paper,
and Allen Lin and Professor John Rickert for
their thoughtful comments on an earlier version of this paper.
D.K.\ would like to thank Davide Gaiotto for suggesting the problem discussed in the second part and to Artem Kalmykov for providing the SAGE code for computing multiplication in the algebra $\mc{A}$. D.K.\ started working on the article when he was a graduate student and then a summer research specialist at MIT.
We thank the anonymous referees for thoughtful comments on this paper.

\pdfbookmark[1]{References}{ref}
\LastPageEnding

\end{document}